\documentclass[seceq,preprint]{ptptex}

\usepackage{graphicx}
\usepackage{wrapft}

\preprintnumber[5cm]{YITP-05-19, KEK-CP-161}

\title{
Low-lying mode contribution to the quenched meson
  correlators in the $\epsilon$-regime
}

\newcommand{\GUAS}{
  School of High Energy Accelerator Science,
  The Graduate University for Advanced Studies (Sokendai),
  Tsukuba 305-0801, Japan
}
\newcommand{\KEK}{
  High Energy Accelerator Research Organization (KEK),
  Tsukuba 305-0801, Japan
}

\author{
Hidenori \textsc{Fukaya}$^{1}$, %
Shoji \textsc{Hashimoto}$^{2,3}$
and Kenji \textsc{Ogawa}$^{3}$
}
\inst{
$^1$Yukawa Institute for Theoretical Physics,
  Kyoto University, Kyoto 606-8502, Japan
\\
$^2$\KEK
\\
$^3$\GUAS
}

\abst{
  We present a quenched calculation of meson correlators 
  with the overlap fermion at very small masses
  2.6--13~MeV. 
  In this region the pion Compton wavelength is larger than
  our lattice size ($L\simeq$ 1.23~fm) and the system is in
  the so-called $\epsilon$-regime of chiral perturbation
  theory. 
  We found that the scalar and pseudo-scalar correlators are
  precisely approximated by a few hundred low-lying fermion
  eigenmodes in this regime, 
  whereas axial-vector correlator receives significant
  contributions from higher eigenmodes.
  We also measure the
  disconnected pseudo-scalar correlator,
  which is well saturated with the low-lying modes.
  Matching these lattice data with the one-loop expressions 
  for the correlators in quenched chiral perturbation
  theory, we evaluate the decay constant 
  $F_\pi$ and the chiral condensate $\Sigma$ as well as the
  parameters $m_0^2$ and $\alpha$, which describe the
  artifacts of the quenched approximation.
}

\begin{document}

\date{\today}
\maketitle

\section{Introduction}\label{sec:intro}

Chiral perturbation theory (ChPT) provides a systematic
method to calculate low energy dynamics of QCD, though it
contains unknown parameters at each order of the expansion
in pion mass squared $m_\pi^2$ and momentum squared
$p_\pi^2$. 
At the lowest order these parameters are the pion decay
constant $F_\pi$ and the chiral condensate $\Sigma$, and there
are 10 other low energy constants in the next-to-leading
order \cite{Gasser:1983yg}.
It is one of the important tasks of lattice QCD to calculate
these low energy constants non-perturbatively from the first 
principles.
In the standard approach, however, this is very demanding
because one has to work on large enough lattices satisfying 
$m_\pi L\gg 1$ to avoid possible finite size effects.
Taking the chiral limit $m_\pi\to 0$ and the continuum limit
while satisfying this condition will be prohibitively 
time consuming even with today's fastest supercomputers,
especially when the quarks are treated dynamically.

In the so-called $\epsilon$-regime 
\cite{Gasser:1987ah,Hansen:1990un,Hansen:1990yg},
where the linear extent of the space-time box is smaller
than the pion Compton wave length $L \ll 1/m_{\pi}$
(but larger than the QCD scale 
$1/\Lambda_{\mathrm{QCD}}\ll L$,
which assures that the pion can be treated as
a point particle and other heavier hadrons are decoupled.
), 
the chiral Lagrangian is still applicable except that the
expansion parameter is given by 
$\epsilon^2\sim m_\pi/\Lambda\sim p_\pi^2/\Lambda^2$, where
$\Lambda$ is a cutoff scale of the chiral Lagrangian roughly
around 1~GeV.
An important observation is that the low energy constants in
the chiral Lagrangian are defined at the cutoff scale
commonly for both the standard and $\epsilon$-regimes.
Therefore, one can determine the low energy constants
in the $\epsilon$-regime and use them in the standard ChPT.
In this way one can avoid the problem of the chiral limit
while keeping the large lattice volume.
Analytic calculation of meson correlation functions in the
$\epsilon$-regime is available for both quenched and
unquenched theories
\cite{Damgaard:2001js,Damgaard:2002qe,Hernandez:2002ds}.

In the study of the chiral regime of lattice QCD
the chiral symmetry plays an essential role.
First of all, one must treat pions near the massless limit, 
which appear as a result of the spontaneous chiral symmetry 
breaking. 
Furthermore, it is known that the effects of fermion zero 
modes become important in the $\epsilon$-regime
\cite{Leutwyler:1992yt} and the correlators depend strongly
on the topological charge of the background gauge field
\cite{Damgaard:2001js,Damgaard:2002qe}.
The lattice fermion formulations preserving chiral symmetry
\cite{Luscher:1998pq} 
(by satisfying the Ginsparg-Wilson relation
\cite{Ginsparg:1981bj}) 
is now commonly used (but only in the quenched approximation).
In this work we use the Neuberger's
overlap-Dirac operator
\cite{Neuberger:1997fp,Neuberger:1998wv}. 
With this formulation there is no fundamental problem to
approach the chiral limit as required in the study of
the $\epsilon$-regime, but the computational cost to invert
the overlap-Dirac operator increases for small quark masses.

In the chiral regime the meson correlators are largely
affected by the low-lying fermion modes
especially by the chiral zero modes.
In this work we explicitly study the effects of such
low-lying modes using the eigenmode decomposition of the
fermion propagator. 
In the quenched approximation we find that the connected
scalar and pseudo-scalar meson correlators are reproduced to
98--99.9\% accuracy (depending on channels) with only 200
lowest-lying eigenmodes on a $10^3\times 20$ at small quark
masses (2.6~MeV $\lesssim m \lesssim$ 13~MeV).
Such saturation was previously found in
Refs.~\citen{DeGrand:2000gq,DeGrand:2003sf}, but it 
should be better
in the $\epsilon$-regime
(In our study, $L$ satisfies
$m_{\pi}L\sim 0.6$ and $\Lambda_{QCD}L\sim 3$.).
An advantage of such eigenmode decomposition is that the
meson correlators can be averaged over space-time points
without much extra computational costs.
The statistical fluctuations originating from the local
bumps of zero-mode wave function are suppressed by the 
space-time averaging and thus we can avoid the large noise
as found in Ref.~\citen{Bietenholz:2003bj}.
The low-mode averaging was also used in recent studies
\cite{DeGrand:2004qw,Giusti:2004yp}.

Matching our numerical data of the axial-vector, scalar and
pseudo-scalar correlators with the quenched ChPT (QChPT)
expressions, 
we extract the leading order low-energy constants $\Sigma$
and $F_\pi$ as well as the parameters appearing due to the
quenched artifact, $\alpha$ and $m_0$.
The axial-vector correlator is most sensitive to $F_\pi$
while $\Sigma$ is precisely determined by the connected
scalar and pseudo-scalar correlators.
We also investigate the chiral condensates and 
the disconnected (hairpin) correlators
for the pseudo-scalar channel.
In general we find good agreement between the lattice data
and the QChPT predictions for topological charge $|Q|$ = 0
and 1 sectors, but the larger topological sectors deviate
significantly, which may suggest breakdown of the
$\epsilon$-expansion in the QChPT at large $|Q|$. 

The axial-vector correlator has already been calculated in
the recent works \cite{Bietenholz:2003bj,Giusti:2004yp}.
Bietenholz {\it et al.} \cite{Bietenholz:2003bj} worked at
a relatively larger quark mass ($\sim$ 21~MeV) and found that
the correlator is well fitted with the QChPT formula for
a large enough lattice size ($L>$ 1.1~fm).
They also pointed out that the signal is very noisy at $|Q|$
= 0.
Using the low-mode averaging technique \cite{DeGrand:2004qw,Giusti:2004yp},
Giusti {\it et al.} pushed the quark mass
down to 10~MeV and found an
encouraging agreement of $F_\pi$ measured in the
$\epsilon$-regime as that from the standard measurement.
Their result is $F_\pi$ = 102(4)~MeV in the
$\epsilon$-regime.
There is another interesting work by the same authors
\cite{Giusti:2003iq}, who investigated the divergent
(as $\sim 1/m^2$) contributions of zero mode in the massless
limit and matched them with the theoretical expectations
from QChPT.

This paper is organized as follows.
In Section~\ref{sec:ChPT}, we review the quenched chiral
perturbation theory (QChPT) in the $\epsilon$-regime
following the discussions of Damgaard {\it et al.}
\cite{Damgaard:2001js}. 
We describe the details of our simulation in
Section~\ref{sec:sim} and study the low-lying eigenmode
dominance in Section~\ref{sec:lowmode}. 
In Section~\ref{sec:meson} 
we present our results for the chiral condensate and meson 
correlators and the comparison with the QChPT.
Conclusions are given in Section~\ref{sec:conclusion}.

\section{Quenched chiral perturbation theory in the 
$\epsilon$-regime}\label{sec:ChPT}

In this section, we briefly review quenched chiral
perturbation theory (QChPT) in the $\epsilon$-regime 
\cite{Damgaard:2001js}
and summarize the relevant formulae for our analysis of
meson correlation functions.

The partition function of QChPT with $N_v$ valence quarks is
written as 
\begin{eqnarray}
  \label{eq:fullpart}
  Z(\theta, M) 
  &=& 
  \int dU \exp \left(
    -\int d^4x \mathcal{L}^{\theta}_M(x)\right),
\end{eqnarray}
where the Lagrangian $\mathcal{L}^{\theta}_M$ is given by
\begin{eqnarray}
  \mathcal{L}^{\theta}_M(x) 
  &=&
  \frac{F_{\pi}^2}{4}\mbox{Str} (\partial_{\mu}U(x)^{-1}\partial_{\mu}U(x))
  -\frac{m\Sigma}{2} \mbox{Str}(U_{\theta}U(x)+U(x)^{-1}U_{\theta}^{-1})
  \nonumber\\
  &&
  +\frac{m_0^2}{2N_c}\Phi(x)^2
  +\frac{\alpha}{2N_c}\partial_{\mu}\Phi(x)\partial_{\mu}\Phi(x)
\end{eqnarray}
at the leading order of $m_\pi^2$ and
$p_\pi^2$ expansion.
The field variable $U(x)$ is integrated over a sub-manifold
of the super-group $Gl(N_v|N_v)$, the maximally symmetric
Riemannian sub-manifold, which is characterized by a matrix of form
\begin{equation}
  U = \left(
    \begin{array}{cc}
      A & B \\
      C & D 
    \end{array}
  \right),
  \;\;\;\;\; A \in U(N_v), 
  \;\;\;\;\; D \in Gl(N_v)/U(N_v),
\end{equation}
and Grassmannian $N_v\times N_v$ matrices $B$ and $C$.
Str denotes the super-trace.
The mass term corresponds to the choice of mass matrix 
$M=(mI_v+m\tilde{I_v})$ with $I_v$ and $\tilde{I_v}$ the
identity matrix in the fermion-fermion and boson-boson
blocks respectively.
The effect of CP violating $\theta$ term enters through
$U_\theta\equiv \exp(i\theta/N_v)I_{N_v}+\tilde{I}_{N_v}$.
In the quenched approximation the singlet field
$\Phi(x)\equiv \frac{F_{\pi}}{\sqrt{2}}\mbox{Str}[-i\ln U(x)]$
does not decouple, and the couplings $m_0^2$ and $\alpha$
are introduced \cite{Sharpe:1992ft}.

The $\epsilon$-regime
\cite{Gasser:1987ah,Hansen:1990un,Hansen:1990yg} 
is realized when the quark mass is
small enough that the pion Compton wavelength $\sim 1/m_\pi$
is larger than the linear extent of the space-time $L$.
The systematic expansion is then reorganized and the
expansion parameter is given by
$\epsilon^2\sim m_\pi/4\pi F_\pi\sim 1/(LF_\pi)^2$.
Unlike the standard ChPT the zero mode of pion gives
important contribution and one must explicitly integrate
out the constant mode of $U(x)$.
This is done by writing as
\begin{equation}
  U(x) = U_0 \exp i\frac{\sqrt{2}\xi(x)}{F_\pi}
\end{equation}
and integrating over the constant mode $U_0$.
One obtains the partition function at a fixed topological
charge $Q$ by Fourier transforming (\ref{eq:fullpart}).
\begin{eqnarray}
  \label{eq:Z_Q}
  Z_{Q}(M) &\equiv& 
  \frac{1}{2\pi}\int_{-\pi}^{+\pi}\!
  d\theta e^{i\theta Q}Z(\theta,M)
  \nonumber\\
  &=&
  \frac{1}{\sqrt{2\pi\langle Q^2\rangle}}e^{-Q^2/2\langle Q^2\rangle}
  \int dU_0d\xi\, (\mathrm{S}\det U_0)^Q \exp\left[
    \frac{m\Sigma V}{2}\mbox{Str}(U_0+U_0^{-1})
  \right.
  \nonumber\\
  &&
  +\left.
    \int\! d^4x \left(
      -\frac{1}{2}\mbox{Str}(\partial_{\mu}\xi\partial_{\mu}\xi)
      -\frac{m_0^2}{2N_c}(\mbox{Str}\xi)^2
      -\frac{\alpha}{2N_c}(\partial_{\mu}\mbox{Str}\xi)^2
    \right)+ O(\epsilon^4)
  \right],
\end{eqnarray}
where $dU_0$ denotes the Haar measure of the maximally Riemannian
sub-manifold of $Gl(N_v | N_v)$.
The topological charge distributes as Gaussian with variance
\begin{equation}
  \label{eq:topsus}
  \frac{\langle Q^2 \rangle}{V} = \frac{F_{\pi}^2m_0^2}{2N_c}
\end{equation}
in the quenched theory, which is an exact equation for $V\to\infty$.
It shows a good contrast with the full theory, for which 
$\langle Q^2 \rangle=m\Sigma V/N_f$ is expected for $N_f$
flavors. 

We note that in the quenched approximation the Gaussian
approximation of the Fourier transform in $\theta$ is
justified only for small topological charge that satisfies
$|Q|\ll\langle Q^2\rangle$.
(See Appendix for the details.)
Therefore, all the results shown below are valid only for
small $|Q|$.
We investigate how this breakdown of the effective theory
occurs using the lattice data.

In the following we consider $N_v$ = 1 and 2 as we are
interested in the system with two light quarks.
All the results are obtained by the perturbation of
$\xi$ fields and the exact integration over zero mode $U_0$,
which can be written in terms of the Bessel functions.

\begin{figure}[tbp]
  \centering
  \includegraphics[width=10cm]{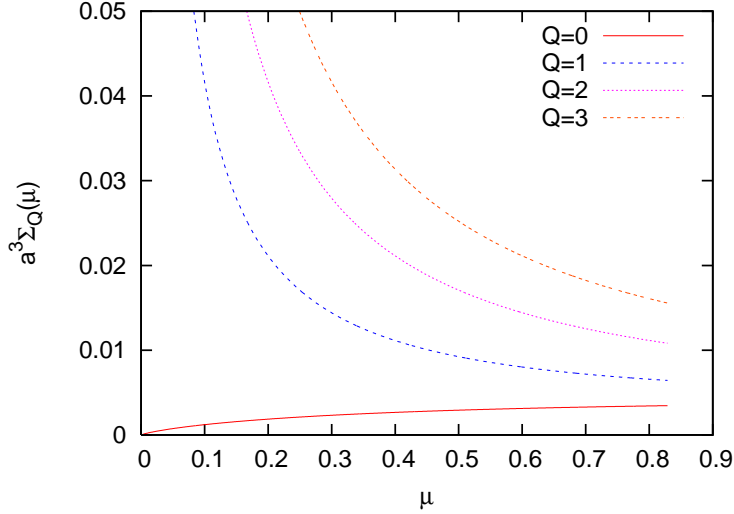}
  \caption{$\Sigma_Q(\mu)$ in different topological sectors at 
$\Sigma^{1/3} =270$MeV.}
  \label{fig:Sigma_Q}
\end{figure}

At the tree-level the scalar condensate is given as
\begin{eqnarray}
  -\langle \bar{\psi}\psi\rangle_Q \equiv 
  \Sigma_Q(\mu) = \Sigma\mu
  (I_{|Q|}(\mu)K_{|Q|}(\mu)+I_{|Q|+1}
  (\mu)K_{|Q|-1}(\mu))
  +\Sigma\frac{|Q|}{\mu}
\end{eqnarray}
with $\mu\equiv m\Sigma V$. 
$I_{|Q|}(\mu)$ and $K_{|Q|}(\mu)$ denote the modified Bessel
functions.
The $\mu$ dependence of $\Sigma_Q(\mu)$ is shown in
Figure~\ref{fig:Sigma_Q}.
Near the massless limit it asymptotically behaves as
$\Sigma_{|Q|}(\mu)\to\Sigma|Q|/\mu$ for $|Q|>0$.
One-loop correction does not change its functional form 
\cite{Damgaard:2001xr}
\begin{eqnarray}
  \label{eq:sigmaeff} 
  \Sigma^{\mathrm{1-loop}}_Q(\mu) 
  = \Sigma_{\mathrm{eff}} \mu^{\prime}
  (I_{|Q|}(\mu^{\prime})K_{|Q|}(\mu^{\prime})+I_{|Q|+1}
  (\mu^{\prime})K_{|Q|-1}(\mu^{\prime}))
  +\Sigma_{\mathrm{eff}}\frac{|Q|}{\mu^{\prime}} 
  =\Sigma_Q(\mu^{\prime}),
\end{eqnarray}
but the parameters $\mu$ and $\Sigma$ are shifted to $\mu'$
and $\Sigma_{\mathrm{eff}}$:
\begin{eqnarray}
  \label{eq:muprime}
  \mu^{\prime} & \equiv & m\Sigma_{\mathrm{eff}}V,
  \\
  \Sigma_{\mathrm{eff}} & \equiv & \Sigma
  \left(
    1+\frac{m_0^2\bar{G}(0)+\alpha\bar{\Delta}(0)}{N_c F_\pi^2}
  \right).
\end{eqnarray}
Here, parameters $\bar{G}(0)$ and $\bar{\Delta}(0)$ are
ultraviolet divergent tadpole integrals,
\begin{eqnarray}
  \bar{G}(x) & \equiv & 
  \frac{1}{V}\sum_{p\neq 0}\frac{e^{ipx}}{p^4},
  \\
  \bar{\Delta}(x) & \equiv &
  \frac{1}{V}\sum_{p\neq 0}\frac{e^{ipx}}{p^2},
\end{eqnarray}
which need to be renormalized.
In our analysis they are to be determined by matching with
lattice data.

Let us define the flavor-singlet meson operators
\begin{eqnarray}
  S^0(x) & \equiv & \bar{\psi}(x)I_{N_V}\psi(x),
  \\
  P^0(x) & \equiv & \bar{\psi}(x)i\gamma_5I_{N_V}\psi(x).
\end{eqnarray}
Adding these operators to the QCD Lagrangian as source terms 
\begin{equation}
  \mathcal{L}\to\mathcal{L}+s(x)S^0(x)+p(x)P^0(x),
\end{equation}
corresponds to a substitution
\begin{equation}
  M \to M+s(x)I_{N_v}+ip(x)I_{N_v}
\end{equation}
in the effective theory. 
The two-point correlation functions of them are obtained by
differentiating the generating functional with respect to
$s(x)$ and $p(x)$. 
To $O(\epsilon^2)$ the results are
\begin{eqnarray}
  \langle S^0(x)S^0(0)\rangle_Q
  &=&
  C_S^0
  +\frac{\Sigma^2}{2F_{\pi}^2}\left[
    \frac{a_-}{N_c}(m_0^2\bar{G}(x)+\alpha\bar{\Delta}(x))
    -\bar{\Delta}(x)\frac{a_++a_--4}{2}
  \right],
  \\
  \langle P^0(x)P^0(0)\rangle_Q
  &=&
  C_P^0
  -\frac{\Sigma^2}{2F_{\pi}^2}\left[
    \frac{a_+}{N_c}(m_0^2\bar{G}(x)+\alpha\bar{\Delta}(x))
    -\bar{\Delta}(x)\frac{a_++a_-+4}{2}
  \right],
\end{eqnarray}
where
\begin{eqnarray}
  a_+ & = &
  4\left[\left(\frac{\Sigma_Q(\mu)}{\Sigma}\right)^{\prime}
    +1+\frac{Q^2}{\mu^2}\right],
  \\
  a_- & = & 
  4\left[-\frac{1}{\mu}\frac{\Sigma_Q(\mu)}
    {\Sigma}+1+\frac{Q^2}{\mu^2}\right],
\end{eqnarray}
and the constant terms are written as
\begin{eqnarray}
  C_S^0 & = &
  \frac{\Sigma^2_{\mathrm{eff}}}{4} a_+^{\mathrm{1-loop}}
  =\Sigma^2_{\mathrm{eff}}\left[
    \left(\frac{\Sigma_Q(\mu')}{\Sigma_{\mathrm{eff}}}\right)^{\prime}
    +1+\frac{Q^2}{\mu'^2}
  \right],
  \\
  C_P^0 & = &
  -\frac{\Sigma^2_{\mathrm{eff}}}{4} a_-^{\mathrm{1-loop}}
  =\Sigma^2_{\mathrm{eff}}\left[
    \frac{1}{\mu'}\frac{\Sigma_Q(\mu')}{\Sigma_{\mathrm{eff}}}
    -\frac{Q^2}{\mu'^2}
  \right].
\end{eqnarray}
Note that the prime denotes the derivative with respect to $\mu$,
\begin{eqnarray}
  \left(\frac{\Sigma_Q(\mu)}{\Sigma}\right)^{\prime}
  =I_{|Q|}(\mu)K_{|Q|}(\mu)-I_{|Q|+1}(\mu)K_{|Q|-1}(\mu)
  -\frac{|Q|}{\mu^2}.
\end{eqnarray}

For flavor non-singlet mesons, we need a $N_v=2$ super-group
integral, which is also described by the Bessel functions.
The non-singlet operators are given by
\begin{eqnarray}
  S^a(x) & \equiv & \bar{\psi}(x) (\tau^a/2) I_{N_V}\psi(x),
  \\
  P^a(x) & \equiv & \bar{\psi}(x) (\tau^a/2) i\gamma_5I_{N_V}\psi(x).
\end{eqnarray}
with the Pauli matrices $\tau^a$.
To $O(\epsilon^2)$ the two-point functions are given by
\begin{eqnarray}
  \label{eq:SS_QChPTconnected}
  \langle S^a(x)S^a(0)\rangle_Q &=& C_S^a
  +\frac{\Sigma^2}{2F_{\pi}^2}\left[
    \frac{c_-}{N_c}(m_0^2\bar{G}(x)+\alpha\bar{\Delta}(x))
    -\bar{\Delta}(x)b_-
  \right],
  \\
  \label{eq:PP_QChPTconnected}
  \langle P^a(x)P^a(0)\rangle_Q &=& C_P^a
  -\frac{\Sigma^2}{2F_{\pi}^2}\left[
    \frac{c_+}{N_c}(m_0^2\bar{G}(x)+\alpha\bar{\Delta}(x))
    -\bar{\Delta}(x)b+
  \right],
\end{eqnarray}
where
\begin{eqnarray}
  b_+ & = & 2\left(1+\frac{Q^2}{\mu^2}\right),
  \\
  b_- & = & 2\frac{Q^2}{\mu^2},
  \\
  c_+ & = & 2\left(\frac{\Sigma_Q(\mu)}{\Sigma}\right)',
  \\
  c_- & =-& 2\frac{1}{\mu}\frac{\Sigma_Q(\mu)}{\Sigma},
\end{eqnarray}
and 
\begin{eqnarray}
  C_S^a & = & 
  \frac{\Sigma^2_{\mathrm{eff}}}{2}\left(\frac{\Sigma_Q(\mu^{\prime})}
    {\Sigma_{\mathrm{eff}}}\right)^{\prime},
  \\
  C_P^a & = & 
  \frac{\Sigma^2_{\mathrm{eff}}}{2}\left(\frac{\Sigma_Q(\mu^{\prime})}
    {\mu'\Sigma_{\mathrm{eff}}}\right).
\end{eqnarray}
For the flavor non-singlet axial-vector current
\begin{equation}
  A_\mu^a(x) = \bar{\psi}(x) (\tau^a/2) i\gamma_\mu\gamma_5 \psi(x)
\end{equation}
the correlator is obtained as \cite{Damgaard:2002qe}
\begin{eqnarray}\label{eq:axial}
  \langle A^a_0(x)A^a_0(0)\rangle_Q
  &=&
  -\frac{F_{\pi}}{V}-2m\Sigma_Q(\mu)\bar{\Delta}(x).
\end{eqnarray}
Here, an important observation is that the axial-current
correlator does not involve the parameters related to the
quenched artifact, {\it i.e.} $m_0^2$ and $\alpha$.
We also note that the constant term is proportional to
$F_\pi$ rather than $\Sigma$.
Therefore, this channel is suitable for an extraction of
$F_\pi$, whereas for the pseudo-scalar and scalar
correlators $F_\pi$ appears only in a coefficient of
$\bar{\Delta}(x)$ and $\bar{G}(x)$ terms.

In the lattice calculation we measure the correlators with
zero spatial momentum projection. 
It is therefore convenient to define
\begin{eqnarray}
  h_1(|t/T|)
  & = &
  \frac{1}{T}\int\! d^3x\, \bar{\Delta}(x)= 
  \frac{1}{2}
  \left[\left(\frac{|t|}{T}-\frac{1}{2}\right)^2
    -\frac{1}{12}\right],
  \\
  h_2(|t/T|)
  & = &
  -\frac{1}{T^3} \int\! d^3x\, \bar{G}(x)
  =\frac{1}{24}\left[\frac{t^2}{T^2}\left(\frac{|t|}{T}-1\right)^2
    -\frac{1}{30}\right].
\end{eqnarray}
In the $\epsilon$-regime the correlators do not behave as
the usual exponential fall-off $\exp(-Mt)$ with the mass gap
$M$ to be expected in the large volume.
Instead, it becomes a simple quadratic function for the
single pole $\bar{\Delta}(x)$ and quartic function for the
double pole $\bar{G}(x)$ integral.

As these expressions show, the meson correlators in the
$\epsilon$-regime are quite sensitive to the topological
charge and the fermion mass. 
Hence they provide a good testing ground for lattice
simulations in the $\epsilon$-regime.
Furthermore the parameters $F_\pi$, $\Sigma$, $m_0$ and
$\alpha$ can be extracted from the fitting of these
correlators.
The parameter $\Sigma$ always appears associated with the
quark mass $m$.
It makes sense because only the combination $m\Sigma$ is
renormalization scale and scheme independent.
The numbers we extract for $\Sigma$ in the following
analysis should be understood as a result in the lattice
regularization at a scale $1/a$.
To relate them with the conventional scheme such as the
$\overline{\mbox{MS}}$ scheme requires perturbative or
non-perturbative matching, which is beyond the scope of this
paper.

\section{Lattice simulations}\label{sec:sim}

We generate the gauge link variables at $\beta=5.85$ in the  
quenched approximation on a $10^3\times 20$ lattice.
The lattice spacing is $a$ = 0.123~fm, which is obtained
from the Sommer scale $r_0$ = 0.5~fm using an interpolation
formula given in Ref.~\citen{Necco:2001xg}.
The linear extent of the lattice is then about 1.23~fm.
We employ the overlap-Dirac operator defined by
\begin{eqnarray}
  D_m & = & \left(1-\frac{\bar{a}m}{2}\right)D + m,
  \\
  D & = & 
  \frac{1}{\bar{a}}\left(1+\gamma_5 \mathrm{sgn} (H_W)\right),
\end{eqnarray}
with the kernel $H_W$ built with the Wilson-Dirac operator
$D_W$, 
\begin{equation}
  H_W=\gamma_5(aD_W-1-s).
\end{equation}
The parameter $s$ controls the negative mass given to $aD_W$
and we choose $s$ = 0.6 at $\beta$ = 5.85 to minimize the
number of low-lying mode in $H_W$.
The symbol sgn denotes a sign function of the large sparse
Hermite matrix $H_W$, and 
$\bar{a}$ is defined as $\bar{a}=a/(1+s)$.
This overlap-Dirac operator satisfies the
Ginsparg-Wilson relation 
\begin{equation}
  \gamma_5D+D\gamma_5 =\bar{a}D\gamma_5D
\end{equation}
exactly, and the $\gamma_5$-hermiticity 
$D^{\dagger}=\gamma_5D\gamma_5$ is also satisfied.
In the practical implementation we approximate the sign
function $\mbox{sign}(H_W)$ using 
the Chebyshev polynomial of degree 100--200 after
subtracting 60 lowest-lying eigenmodes of $\mbox{sign}(H_W)$
exactly. 
The error of the sign function is then 10$^{-12}$ level and
safely neglected in our numerical results.

\begin{figure}[hbtp]
\begin{center}
\includegraphics[width=10cm]{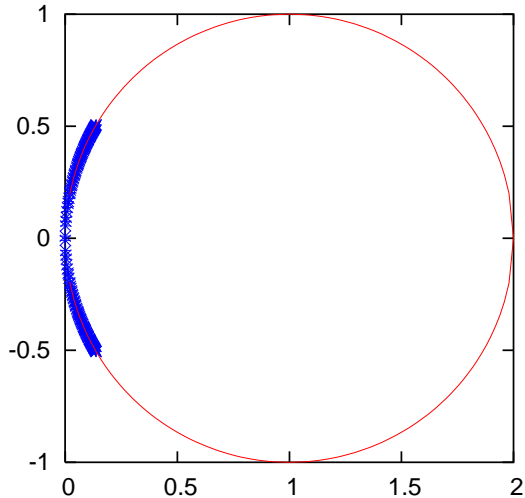}
\caption{
  Lowest 202 eigenvalues of the overlap-Dirac operator 
  at beta=5.85 on a $10^320$ lattice of
  topological charge $Q=-2$.
  The eigenvalues cover a $\pi/3$ arc of the circle. 
}\label{fig:eigenV}
\end{center}
\end{figure}

One of the essential points of our work is to use the
eigenmode decomposition of the fermion propagator.
For this purpose we calculate 100 lowest (but non-zero)
eigenvalues of $P_{\pm}DP_{\pm}$ and their eigenfunctions as
well as zero-mode eigenfunctions of $P_{\mp}DP_{\mp}$ 
for negative (and positive) topological charge.
We use the numerical package ARPACK \cite{ARPACK}, which
implements the implicit restarted Arnoldi method.
The chiral projection operator 
$P_{\pm}\equiv (1\pm\gamma_5)/2$
is applied in order to reduce the rank of the matrix.
The eigenvalues and eigenvectors of the original matrix $D$
can be reconstructed from those for the chirally projected
operators.
We thus obtain 200+$|Q|$ eigenmodes of $D$ for each gauge
configuration. 
Note that these 200+$|Q|$ eigenvalues cover more than 15\% 
of the circle in the complex space of the eigenvalues of $D$
as Figure~\ref{fig:eigenV} shows. 
The topological charge is obtained from the number of 
zero-modes and their chirality.
The number of configurations for each topological
sector is given in Table~\ref{tab:confnum}.
We analyze the gauge configurations of $|Q|\leq 3$ sectors. 

\begin{table}[btp]
\begin{center}
\caption{
  Number of configurations in each topological sector.
}\begin{tabular}{ccccc}
\hline\hline
$|Q|$ & 0 & 1 & 2 & 3 \\
\hline
\# of confs. & 20 & 45 & 44 & 24 \\
\hline
\end{tabular}\label{tab:confnum}
\end{center}
\end{table}

When the exact inverse of the overlap-Dirac operator is
needed, we use the techniques described in
Ref.~\citen{Giusti:2002sm}. 
For a given source vector $\eta$, we solve the equation 
\begin{equation}
  D_m\psi=\eta
\end{equation}
by separating the left and right handed components as
$\psi = P_- \psi + P_+ \psi$ and solving two equations
\begin{eqnarray}
  P_-\psi &=& (P_-D^{\dagger}_mD_mP_-)^{-1}P_-D^{\dagger}_m\eta,
  \\
  P_+\psi &=& (P_+D_mP_+)^{-1}(P_+\eta-P_+D_mP_-\psi),
\end{eqnarray}
consecutively.
(The above equations apply for positive $Q$ and the same
procedure applies with a replacement $P_+\leftrightarrow
P_-$ for negative $Q$.)
We use the conjugate gradient (CG) algorithm to invert
the chirally projected matrices with the low-mode
preconditioning with 20 lowest eigenmodes. 
With this low-mode preconditioning we gain about one order
of magnitude speed-up of the CG solver
(we need only 20-40 iterations for each $P_{\pm}\psi$.)
for our smallest
quark mass 0.0016, which corresponds to 2.6~MeV in the
physical unit.


\section{Low-mode dominance for the meson correlators}
\label{sec:lowmode}

The inverse of the overlap-Dirac operator $D$ can be
decomposed into the contributions from each eigenmode with
eigenvalue $\lambda_i$ and eigenvector $v_i(x)$ as
\begin{equation}
  \label{eq:eigen_decomp}
  D^{-1}_m(x,y) =
  \sum_{i=1}^{N_{\mathrm{low}}}
  \frac{1}{(1-\bar{a}m/2) \lambda_i+m} v_i(x)v_i^{\dagger}(y)
  +\Delta D^{-1}_m(x,y).
\end{equation}
Here, the eigenmode decomposition is done incompletely, and
the sum is truncated at some cutoff $N_{\mathrm{low}}$, which
we set $N_{\mbox{\tiny low}}=200+|Q|$.
The additional term $\Delta D^{-1}_m(x,y)$ represents the
contribution from higher eigenmodes. 

We expect that the low energy physics is dominated by 
the low-lying eigenmodes.
Near the massless limit the lowest-lying eigenmodes play a
dominant role as they are enhanced by a $1/\lambda_i$
factor, and especially the zero modes give divergent
contribution in the quenched approximation.
(In the unquenched case their occurrence is suppressed by the
fermion determinant.)
Sensitivity to the gauge field topology mostly comes from
these low-lying eigenmodes.
Therefore, the eigenmode decomposition
(\ref{eq:eigen_decomp}) is expected to give a good
approximation for low energy physics even if we ignore the
higher mode contribution $\Delta D^{-1}_m(x,y)$.
The cutoff $N_{\mathrm{low}}$ must be large enough for such 
approximation to cover the all relevant low-lying modes, and  
it depends on the pion mass and physical volume of the
system. 
For massive fermions the eigenmodes below
$\lambda_i\lesssim m$ become equally important, and 
we need larger $N_{\mathrm{low}}$ than in the massless
limit.

On the other hand, the short distance physics could be
affected by the higher eigenmodes.
It can be seen by looking at the defining equation
$\sum_z D_m(x,z) D_m^{-1}(z,y) = 
\sum_{i=1}^{N_{\mathrm{low}}} v_i(x)v_i^\dagger(y) +
D_m(x,z) \Delta D_m^{-1}(z,y) = \delta(x,y)$
at $x=y$.
Since the sum
$\sum_{i=1}^{N_{\mathrm{low}}} v_i(x)v_i^\dagger(x)$
approaches monotonically to one when 
$N_{\mathrm{low}}$ is sent to the size of the matrix
$N_{\mathrm{max}}$, the contribution from the remaining term  
$\Delta D_m^{-1}(x,y)$ becomes significant for 
$N_{\mathrm{low}}$ much smaller than $N_{\mathrm{max}}$.
However, such a short distance correlation described by 
$\Delta D_m^{-1}(x,y)$ should be insensitive to the gauge
field topology.

\subsection{Connected correlators}
First, let us consider the ``connected'' meson correlators
\begin{eqnarray}
  M_i(t)M_i(0)_{\mathrm{conn}} &\equiv& 
  \sum_{\vec{x}}\bar{\psi}\Gamma_i\psi(\vec{x},t) 
  \bar{\psi}\Gamma_i\psi(\vec{0},0)
  _{\mathrm{conn}}
  \nonumber\\
  &=&
  -\sum_{\vec{x}}\mbox{tr}\left(
    \Gamma_i D_m^{-1}(\vec{x},t; \vec{0},0)
    \Gamma_i D_m^{-1}(\vec{0},0; \vec{x},t)
  \right),
\end{eqnarray}
where $M_i$ denotes local operator corresponding to 
pseudo-scalar (PS), scalar (S), vector (V) and axial-vector
(AV) currents, and
$\Gamma_i$ denotes the corresponding gamma matrix,
$\Gamma_{\mathrm{PS}}=i\gamma_5$,
$\Gamma_{\mathrm{S}}=1$,
$\Gamma_{\mathrm{V}}=i\gamma_0$, and
$\Gamma_{\mathrm{AV}}=i\gamma_5\gamma_0$.

\begin{figure}[tbp]
  \begin{center}
    \includegraphics[width=8cm]{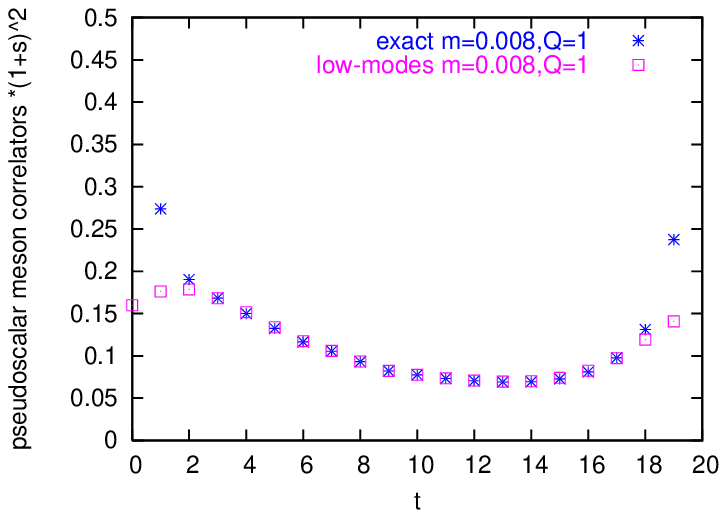}
    \includegraphics[width=8cm]{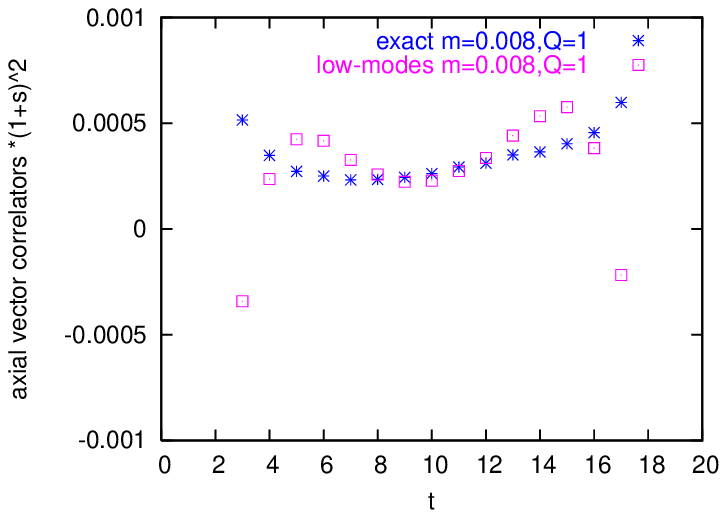}
    \caption{
      Pseudo-scalar (left) and axial-vector (right)
      correlators at $m$ = 0.008 and $|Q|=1$.
      The low-mode-approximated correlator (\ref{eq:lowapp})
      is compared with the corresponding exact one.
    }\label{fig:accuracy}
  \end{center}
\end{figure}

\begin{figure}[tbp]
  \begin{center}
    \includegraphics[width=8cm]{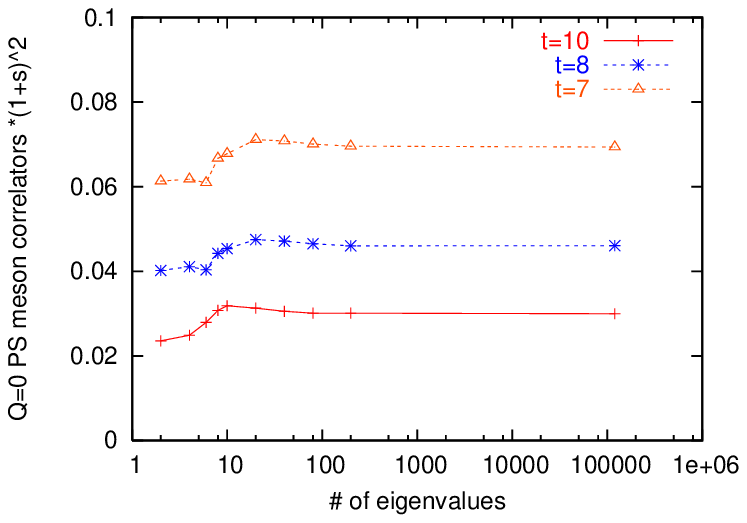}
    \includegraphics[width=8cm]{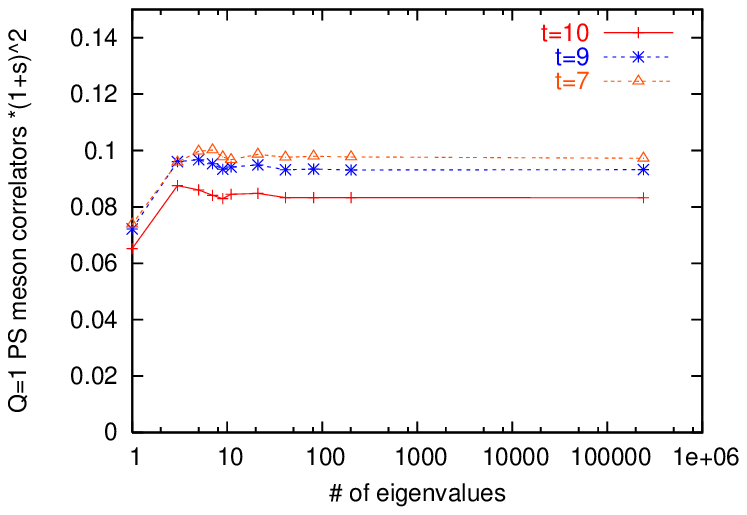}
    \caption{
      Saturation of the connected pseudo-scalar correlator
      for one sample configuration at $m=0.008$.
      Two plots are shown for $|Q|=0$ (left) and $|Q|=1$ (right). 
      The rightmost points around 120,000 correspond to the
      exact correlator obtained by the CG method.
    }
    \label{fig:saturation}
  \end{center}
\end{figure}

We calculate these correlators in two ways: one is an
exact calculation with the conjugate gradient (CG) method
and the other is the low-mode approximation
\begin{equation}
  \label{eq:lowapp}
  D^{-1}_m(x,y)\sim\sum_{i=1}^{200+|Q|}
  \frac{1}{(1-\bar{a}m/2)\lambda_i+m}v_i(x)v_i^{\dagger}(y),
\end{equation} 
where the higher eigenmode contributions are neglected.
Comparison of them at $m$ = 0.0016, 0.0048, 0.008 
($\lesssim$ 13~MeV)
is shown in Table~\ref{tab:accuracy}, where
maximal numerical difference between the two correlators in
the region $7\leq t \leq 13$ is listed for each operator and
topological sector.
For the scalar and pseudo-scalar mesons, 
the approximation (\ref{eq:lowapp}) does work well to
98--99.9\% accuracy,
while the lowest $200+|Q|$ modes are not enough to reproduce
the vector and axial vector correlators. 
Figure~\ref{fig:accuracy} shows the pseudo-scalar and
axial-vector correlators at $m$ = 0.008 and $|Q|=1$.
We observe very good agreement for the pseudo-scalar
correlator for wide region of $t$ (from 3 to 17), but the
axial-vector case is much worse.
This is probably because the axial-vector correlator is
smaller in magnitude by a factor of $O(m)$, and therefore
small fluctuation of eigenvectors is enhanced
\cite{Blum:2000kn}.
Figure~\ref{fig:saturation} shows how the low-lying mode
contributions saturate to the full correlators.
We find that about 100 lowest eigenmodes suffice to
approximate the full correlator to a very good accuracy.
The plot is shown for $m=0.008$, but the saturation becomes
even better for smaller quark masses.

\begin{table}[tbp]
\begin{center}
\caption{
  Comparison of the low-mode approximated ``connected''
  correlators with their exact ones.
  The approximation uses the quark propagator
  (\ref{eq:lowapp}) with $200+|Q|$ low-lying modes
  at $m$ = 0.0016, 0.0048, 0.008.
  The maximum deviation in $7\leq |t| \leq 13$ is shown.
  The number of configurations is given in
  Table~\ref{tab:confnum}. 
}
\begin{tabular}{ccccc}
\hline
\hline
$m=0.008$ ($\sim$ 13~MeV) 
& $|Q|=0$ & $|Q|=1$ &$|Q|=2$ &$|Q|=3$ \\
\hline
 scalar        & 2.01\%  & 1.49\%  & 0.46\% & 0.44\% \\
 pseudo-scalar & 1.19\%  & 0.48\%  & 0.28\% & 0.22\% \\
 vector        & 41.6\%  & 258 \%  & 259\%  & 94.6\%   \\
 axial-vector  & 28.4\%  & 40.5\% &  24.3\%  & 24.2\% \\
\hline
$m=0.0048$ ($\sim$ 7.7~MeV)
& $|Q|=0$ & $|Q|=1$ &$|Q|=2$ &$|Q|=3$ \\
\hline
 scalar        & 1.57\%  & 0.65\%  & 0.16\% & 0.18\% \\
 pseudo-scalar & 1.06\%  & 0.33\%  & 0.10\% & 0.08\% \\
 vector        & 28.9\%  & 198 \%  & 110\%  & 104\%   \\
 axial-vector  & 26.2\%  & 38.6\% &  19.7\%  & 18.3\% \\
\hline
$m=0.0016$ ($\sim$ 2.6~MeV)
& $|Q|=0$ & $|Q|=1$ &$|Q|=2$ &$|Q|=3$ \\
\hline
 scalar        & 1.41\%  & 0.08\%  & 0.03\% & 0.03\% \\
 pseudo-scalar & 0.95\%  & 0.06\%  & 0.04\% & 0.02\% \\
 vector        & 22.5\%  & 149 \%  & 157\%  & 166\%   \\
 axial-vector  & 21.5\%  & 32.6\% &  20.4\%  &34.7\% \\
\hline
\end{tabular}\label{tab:accuracy}
\end{center}
\end{table}

\begin{figure}[tbp]
\begin{center}
\includegraphics[width=8.7cm]{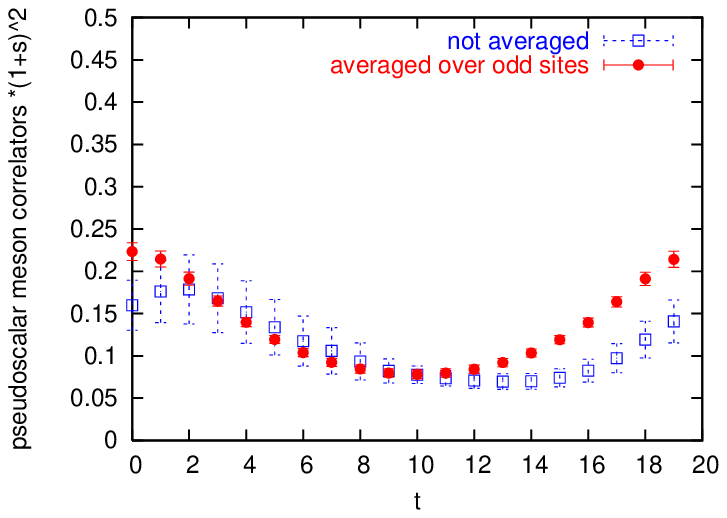}
\caption{
  The pseudo-scalar correlator at $m=0.008$ and $|Q|=1$.
  Filled symbols denote the data with the low-mode
  averaging, while open symbols are not averaged.
}\label{fig:LMA}
\end{center}
\end{figure}

An advantage of the low-mode approximation (\ref{eq:lowapp})
is that $D^{-1}_m(x,y)$ at any $x$ and $y$ is obtained
without performing the CG inversion, so that one can easily
average the source point over the space-time
\begin{equation}
  \left\langle
    \sum_{\vec{x}} \bar{\psi}\Gamma_i\psi(\vec{x},t) 
    \bar{\psi}\Gamma_i\psi(\vec{0},0)
  \right\rangle^Q_{\mathrm{conn}}
  \to \frac{1}{T L^3}
  \sum_{\vec{x}_0,t_0}
  \left\langle 
    \sum_{\vec{x}}\bar{\psi}\Gamma_i\psi(\vec{x},t+t_0) 
    \bar{\psi}\Gamma_i\psi(\vec{x}_0,t_0)
  \right\rangle^Q_{\mathrm{conn}}.
\end{equation}
This so-called low-mode averaging dramatically reduces 
the fluctuation of the low-lying modes as shown in
Figure~\ref{fig:LMA}, which is also reported in
Ref.~\citen{Giusti:2004yp}. 
In practice we average only over $(L/2)^3\times(T/2)$
lattice points where the site index is an even number for
each direction. 
Note that even after the low-mode averaging,
the error from the truncation of the higher modes
is negligible compared to the statistical error 
$\sim$ 15\% in $Q=0$ sector 
and $\sim$ 5\% in $Q\neq 0$ sectors in the range
$7\leq t\leq 13$.

\subsection{Chiral condensates}

\begin{figure}[tbp]
\begin{center}
\includegraphics[width=8cm]{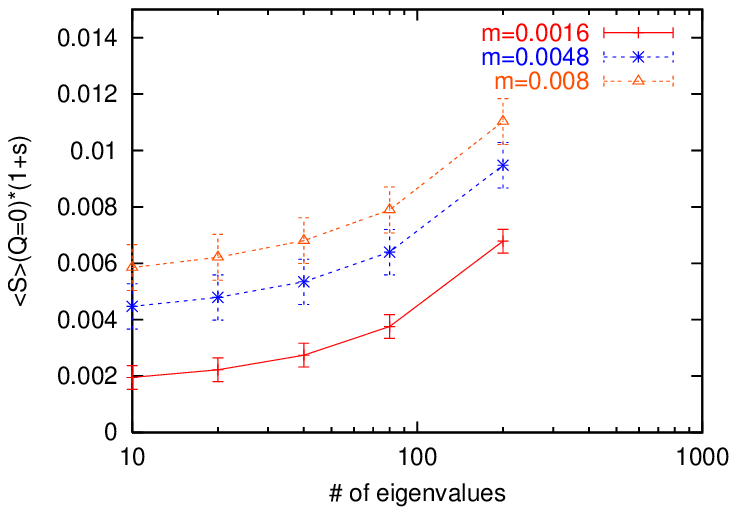}
\includegraphics[width=8cm]{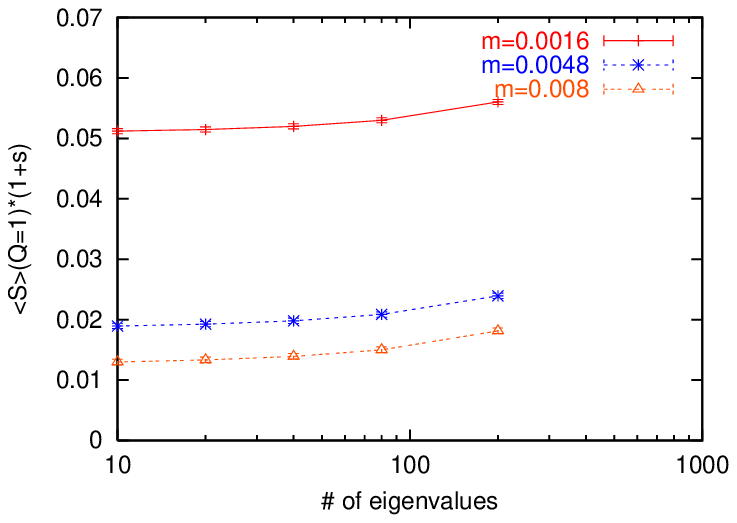}
\caption{
  Low-mode saturation for $-\langle \bar{\psi}\psi\rangle^Q$.
  Data are averaged over 20 gauge configurations of 
  $Q=0$ (left) and $Q=1$ (right).
}\label{fig:saturationconden1}
\end{center}
\end{figure}

\begin{figure}[tbp]
\begin{center}
\includegraphics[width=8cm]{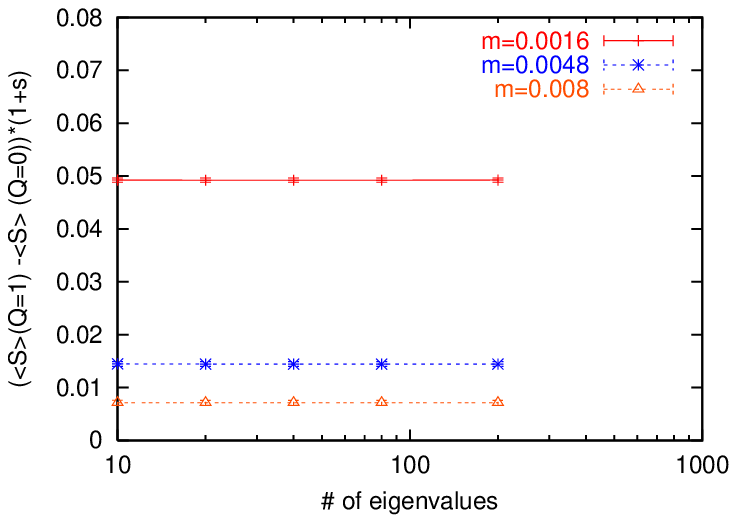}
\includegraphics[width=8cm]{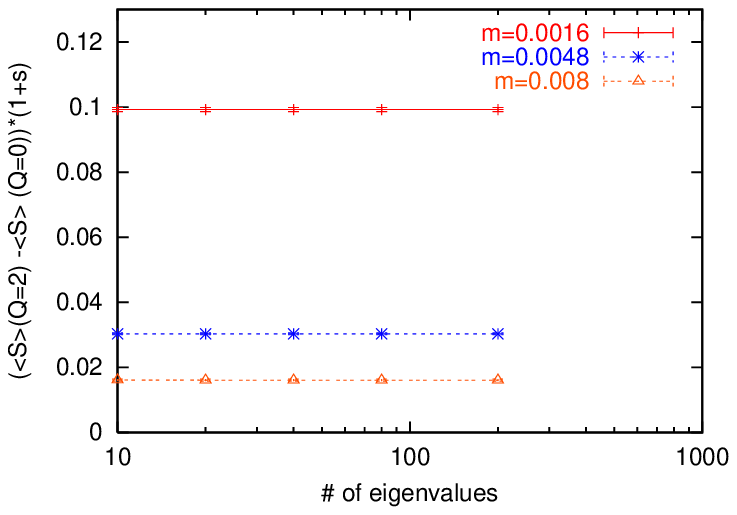}
\caption{
  Low-mode saturation for 
  $ -(\langle \bar{\psi}\psi\rangle^Q - \langle \bar{\psi}\psi\rangle^0)$
averaged over 20 configurations at  $Q=1$ (left) and $Q=2$ (right).
}\label{fig:saturationconden2}
\end{center}
\end{figure}

Consider the low-mode contribution to the scalar and
pseudo-scalar condensates for a fixed topological charge $Q$ 
\begin{eqnarray}
  \langle \bar{\psi}\psi(x)\rangle^Q 
  &=&
  - \langle \mbox{tr}D_m^{-1}(x,x)\rangle^Q
  \nonumber\\
  &=& 
  -\left\langle \mbox{tr}\left(
      \sum_{i=1}^{N_{\mbox{\tiny low}}}
      \frac{1}{(1-\bar{a}m/2)\lambda_i+m}v_i(x)v_i^{\dagger}(x)
      +\Delta D_m^{-1}(x,x)
    \right)\right\rangle^Q,
  \\
  \langle \bar{\psi}\gamma_5\psi(x)\rangle^Q 
  &=&
  - \langle\mbox{tr}\gamma_5D_m^{-1}(x,x)\rangle^Q
  \nonumber\\
  &=&
  -\left\langle\mbox{tr}\left(
      \sum_{i=1}^{N_{\mbox{\tiny low}}}\frac{1}
      {(1-\bar{a}m/2)\lambda_i+m}\gamma_5v_i(x)v_i^{\dagger}(x)
      +\gamma_5\Delta D_m^{-1}(x,x)
    \right)\right\rangle^Q.
\end{eqnarray}
For the scalar condensates, it is known that 
$\Delta D_m^{-1}(x,x)$ includes unwanted additive 
ultraviolet divergences \cite{Hernandez:1999cu}
\begin{equation}\label{eq:div}
  \langle \mbox{tr}D_m^{-1}(x,x)\rangle^Q
  = \frac{6}{(1+s)a^3}+C_2\frac{m}{a^2}+C_1\frac{m^2}{a}
  +\Sigma_Q(\mu^{\prime})+O(\epsilon^4),
\end{equation}
where $C_2$ and $C_1$ are unknown constants.
The first term comes from the modification of the chiral
symmetry in the Ginsparg-Wilson relation
\cite{Ginsparg:1981bj}, 
$\gamma_5 D^{-1}(x,y)+D^{-1}(x,y)\gamma_5=\bar{a}\delta_{x,y}$,
at $x=y$.
If $N_{\mathrm{low}}$ is large enough,
the higher mode contribution $\Delta D_m^{-1}(x,x)$ should
be insensitive to the link variables $U_{\mu}(y)$ separated
large enough from $x$ and thus to the global structure of
the gauge field configuration, such as the topological
charge. 
We, therefore, expect that such contribution vanishes in the
difference between different topological sectors,
\begin{equation}
  -(\langle \bar{\psi}\psi(x)\rangle^Q 
  - \langle \bar{\psi}\psi(x)\rangle^0).
\end{equation}
In other words this difference should be well described only
by the low-lying eigenmodes.
In fact we observe such a low-mode saturation as
shown in Figures~\ref{fig:saturationconden1} and  
\ref{fig:saturationconden2}, while the individual condensate
$-\langle \bar{\psi}\psi(x)\rangle^Q$ is not saturated.

\begin{figure}[tbp]
\begin{center}
\includegraphics[width=8cm]{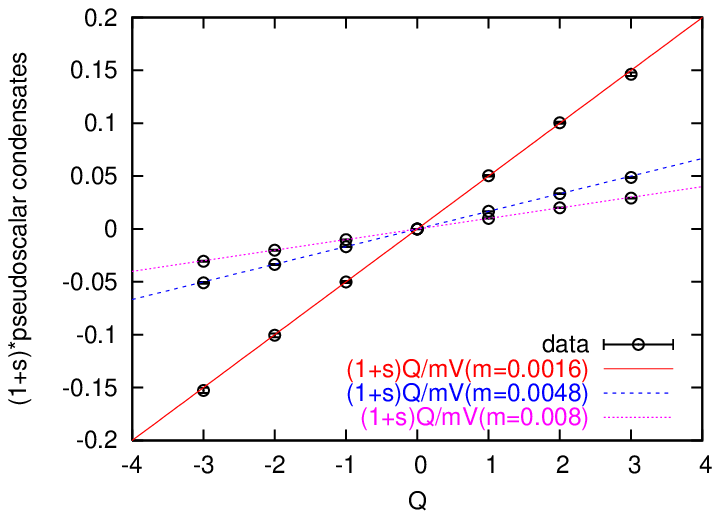}
\caption{
  Pseudo-scalar condensates approximated with 
  $N_{\mbox{\tiny low}}=200+|Q|$
  low-modes.
  The lines represent the expectation (\ref{eq:pscond}).
}\label{fig:psconden}
\end{center}
\end{figure}

For the pseudo-scalar condensate, we do not have to take
care of the higher modes because the condensate is
determined only by the zero modes and the contributions from
other eigenmodes cancel because of the orthogonality among
different eigenvectors. 
As shown in Figure~\ref{fig:psconden}, our data with
$N_{\mbox{\tiny low}}=200+|Q|$ low-modes perfectly agree
with the theoretical expectation
\begin{equation}\label{eq:pscond}
  - \langle \bar{\psi}\gamma_5\psi\rangle_Q = \frac{Q}{mV}.
\end{equation}

\subsection{Disconnected correlators}
\label{sec:low-mode-disconnected}
Here we discuss the expectation value of the
``disconnected'' diagrams for a fixed topological charge. 
As in the case of the connected diagram, we expect
\begin{eqnarray}
  \langle M_i(t)M_i(0)\rangle_{\mathrm{disc}}^Q 
  & \equiv & 
  \left\langle
    \sum_{\vec{x}}
    \bar{\psi}\Gamma_i\psi(\vec{x},t) 
    \bar{\psi}\Gamma_i\psi(\vec{0},0)
  \right\rangle_{\mathrm{disc}}^Q
  \nonumber\\
  & = &
  \left\langle
    \sum_{\vec{x}} \mbox{tr}\left(
      \Gamma_i D_m^{-1}(\vec{x},t; \vec{x},t)
    \right)
    \mbox{tr}\left(
      \Gamma_i D_m^{-1}(\vec{0},0; \vec{0},0)
    \right)
  \right\rangle^Q,
  \nonumber\\
  & \stackrel{t\gg0}{\longrightarrow} & 
  \left\langle
    \sum_{\vec{x}}\mbox{tr}\left(
      \Gamma_i 
      \sum_{i=1}^{200+|Q|}
      \frac{1}{(1-\bar{a}m/2)\lambda_i+m}
      v_i(x)v_i^{\dagger}(x)
    \right)\right.
  \nonumber\\
  &&
  \left.
    \times\mbox{tr}\left(
      \Gamma_i \sum_{i=1}^{200+|Q|}
      \frac{1}{(1-\bar{a}m/2)\lambda_i+m}
      v_i(0)v_i^{\dagger}(0)
    \right)
  \right\rangle^Q
  \nonumber\\
  & + &
  2 \left\langle\mbox{tr}
    \left(\Gamma_i\Delta D_m^{-1}(x,x)\right)
  \right\rangle^{\prime}
  \nonumber\\
  &&
  \times \left\langle \mbox{tr}
    \left(\Gamma_i
      \sum_{i=1}^{200+|Q|}
      \frac{1}{(1-\bar{a}m/2)\lambda_i+m}v_i(0)v_i^{\dagger}(0)
    \right)
  \right\rangle^Q
  \nonumber\\
  & + &
  \left[
    \left\langle \mbox{tr}
      \left(\Gamma_i\Delta D_m^{-1}(x,x)\right)
    \right\rangle^{\prime}
  \right]^2,
\end{eqnarray}
where $x=(\vec{x},t)$.
We assume that higher modes' contribution does not have
correlation with any local operator $O(y)$ separated enough
from $x$, {\it i.e.}
\begin{eqnarray}
  \langle \Delta D_m^{-1}(x,x) O(y) \rangle^Q
  \stackrel{|x-y|\gg0}{\longrightarrow}  
  \langle \Delta D_m^{-1}(x,x) \rangle'
  \times \langle O(y) \rangle^Q,
\end{eqnarray}
where the expectation value $\langle\cdots\rangle'$
represents insensitivity to the topological charge.
We also use the translational invariance
$\langle O(x)\rangle=\langle O(0)\rangle$.
Unlike in the ``connected'' case, we cannot check the
low-mode dominance by explicitly computing the exact
correlators because the numerical cost is too expensive.
However, for the pseudo-scalar disconnected diagram 
$\langle \mbox{tr}\gamma_5\Delta D_m^{-1}(x,x)\rangle^{\prime}$
vanishes because $\Delta D_m^{-1}(x,x)$ does not contain the
zero modes.

\begin{figure}[tbp]
\begin{center}
\includegraphics[width=8cm]{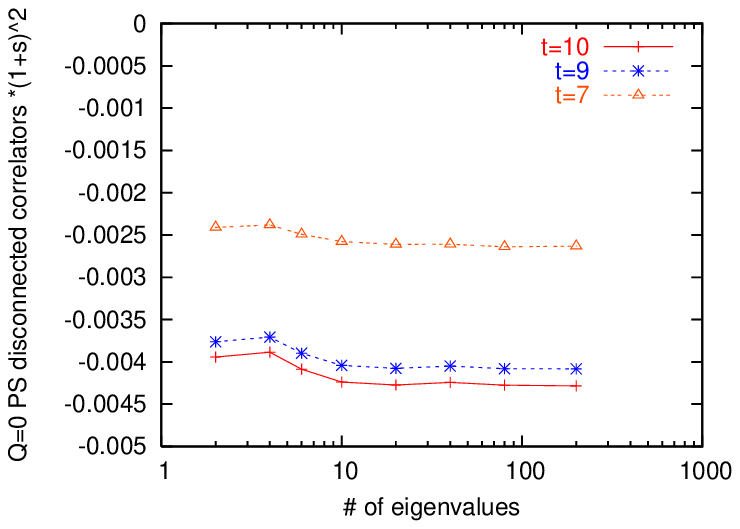}
\includegraphics[width=8cm]{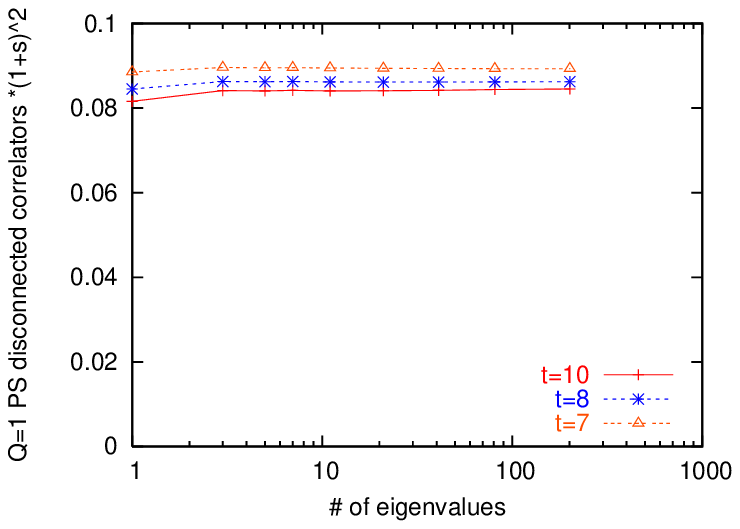}
\caption{
  Saturation of the pseudo-scalar disconnected correlators
  for one sample configuration
  at $m=0.008$ for $Q=0$ (left) and $Q=1$ (right).
}\label{fig:saturationdis}
\end{center}
\end{figure}

In fact, as Figure~\ref{fig:saturationdis} shows, we find 
good saturation with the lowest 200 eigenmodes 
for the pseudo-scalar disconnected correlators.
Similar results were also obtained previously
in the study of the $\eta'$ propagator with the Wilson fermion
\cite{Neff:2001zr} and with the overlap fermion \cite{DeGrand:2002gm}.

\section{Extraction of the low energy constants from the
  meson correlators}
\label{sec:meson}

\subsection{$F_{\pi}$ from the axial-vector
  correlator} 
The axial-vector current correlator (\ref{eq:axial}) is most
sensitive to $F_\pi$ and not contaminated by the parameters
$m_0$ and $\alpha$.
The problem is, however, that the low-mode
dominance does not hold, and we have to solve the quark
propagators exactly.
Hence, the statistical noise could become a problem as we
cannot average the source point over space-time.
Our strategy is to treat data at different quark 
masses and different topologies at the same time 
to reduce such large statistical errors.

We use the local axial-current 
$A_{\mu}^a(x)=\bar{\psi}(x)\gamma_5\gamma_{\mu}(\tau^a/2)\psi(x)$,
constructed from the overlap fermion field $\psi(x)$.
Since it is not the conserved current corresponding to the
lattice chiral symmetry, (finite) renormalization is needed
to relate it to the continuum axial-vector current.
To calculate the $Z_A$ factor non-perturbatively, we follow
the method applied in Refs.~\citen{Giusti:2001yw,Giusti:2001pk}.
Namely, we calculate
\begin{eqnarray}
  \label{eq:R_rho}
  aR_{\rho}(t) \equiv 
  \frac{
    a\sum_{\vec{x}}
    \langle \bar{\nabla}_0 A^a_0(\vec{x},t)P^a(0,0)\rangle
  }{
    \sum_{\vec{x}}\langle P^a(\vec{x},t)P^a(0,0)\rangle
  },
\end{eqnarray}
where $\bar{\nabla}_0$ denotes a symmetric lattice
derivative.
The pseudo-scalar density $P^a(x)$ must be the chirally
improved one
$\bar{\psi}(\tau^a/2)\gamma_5(1-\frac{\bar{a}}{2}D)\psi$
associated with the exact chiral symmetry on the lattice.
For the on-shell matrix elements such as the one considered
here, one can use the equation of motion to replace the
$\bar{a}D$ term by $-am/(1-\bar{a}m/2)$, which is negligible
for our quark masses.
We therefore use the local operator for $P^a(x)$.

\begin{figure}[tbp]
\begin{center}
\includegraphics[width=8cm]{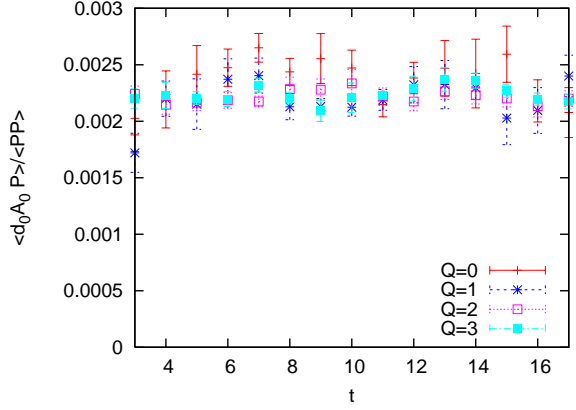}
\caption{
  $aR_{\rho}(t)$ at $m=0.0016$.
  Data for topological sectors $|Q|$ = 0, 1, 2, and 3 are
  shown. 
}\label{fig:AWIQ}
\end{center}
\end{figure}

\begin{figure}[tbp]
\begin{center}
\includegraphics[width=8cm]{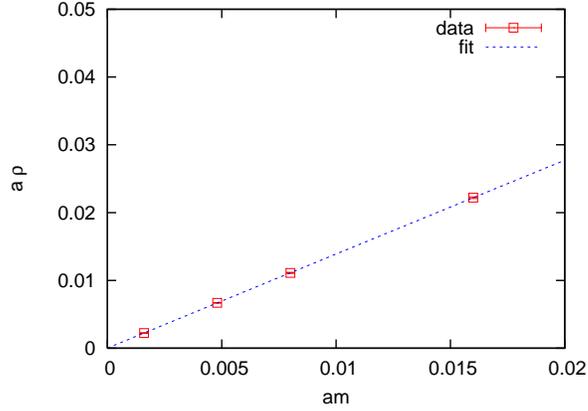}
\caption{
  $a\rho$ as a function of the bare quark mass. 
  Dashed line is a result of the quadratic fit:
  $a\rho=0.00001+1.390(ma)-0.19(ma)^2$. 
}\label{fig:ZAfit}
\end{center}
\end{figure}

The ratio (\ref{eq:R_rho}) turns out to be insensitive to
topology as shown in Figure~\ref{fig:AWIQ}.
Fitting the average of $R_{\rho}(t)$ over all topological
sectors with a constant in the range $7\leq t \leq 13$,
we obtain the results for
\begin{equation}
  a\rho(ma)
  \equiv
  \frac{a
    \langle \bar{\nabla_{\mu}}A^a_{\mu}(x)P^a(0)\rangle
  }{
    \langle P^a(x)P^a(0)\rangle
  }
  =
  \frac{2ma}{Z_A}+O(a^2)
\end{equation}
at four quark masses 
$m$ = 0.0016, 0.0048, 0.008, 0.0160,
which are shown in Figure~\ref{fig:ZAfit}.
With a quadratic fit we obtain 
\begin{equation}
  a\rho(ma)=0.00001(2)+1.390(14)(ma)-0.19(74)(ma)^2.
\end{equation}
The constant term is perfectly consistent with zero and we
can extract $Z_A$ from the linear term as $Z_A$ = 1.439(15)
, which is consistent with the value $Z_A$ = 1.448(4) 
reported in Ref.~\citen{Bietenholz:2004wv} which was done
with the same $\beta$ and $s$.

\begin{figure}[tbp]
\begin{center}
\includegraphics[width=8cm]{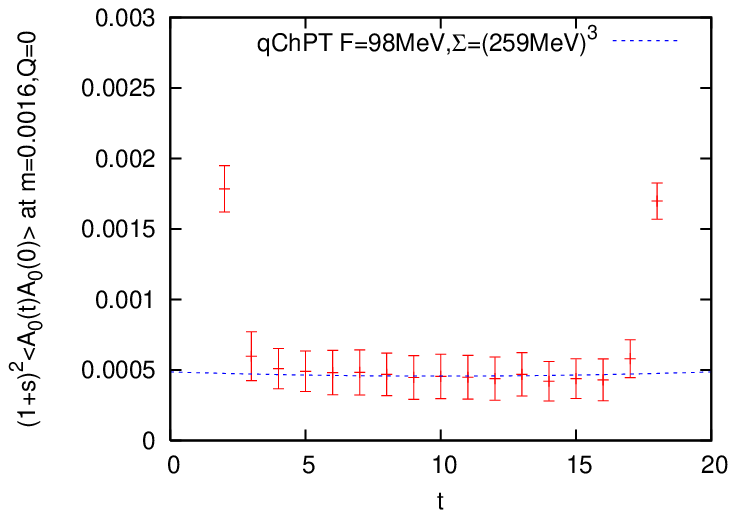}
\includegraphics[width=8cm]{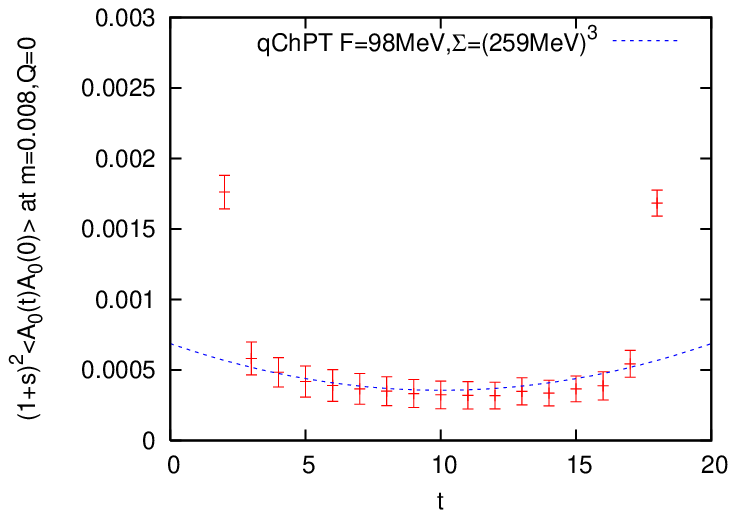}
\includegraphics[width=8cm]{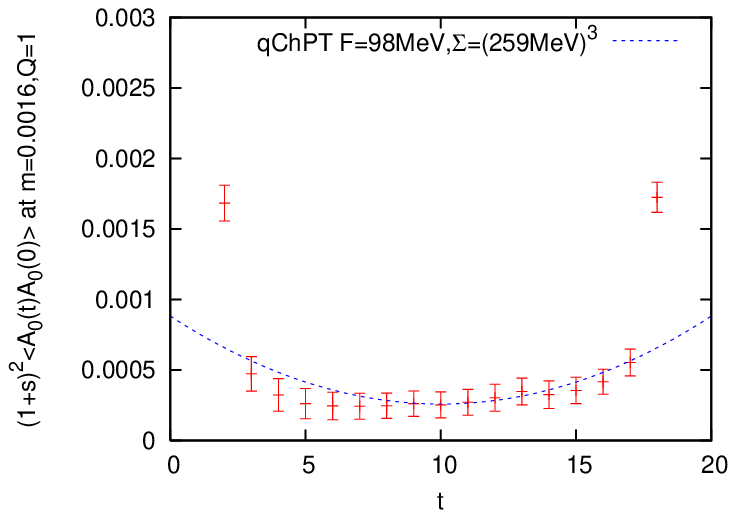}
\includegraphics[width=8cm]{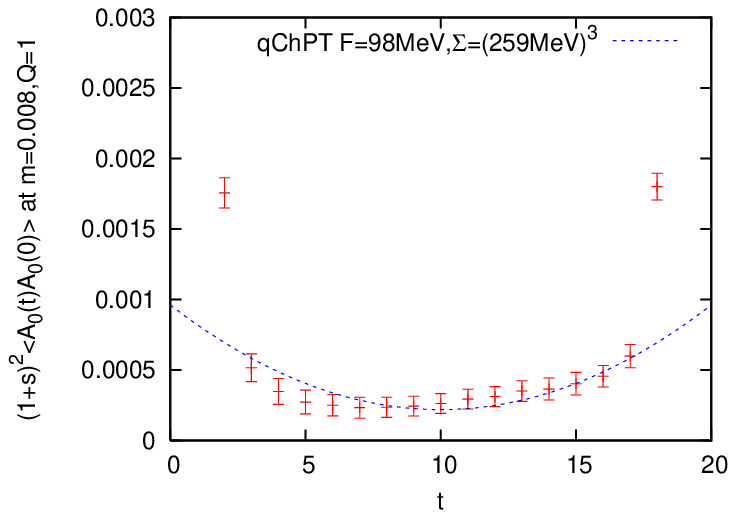}
\caption{
  Axial-vector current correlators at $m$ = 0.0016 (left)
  and 0.008 (right) for $|Q|$ = 0 (top) and 1 (bottom).
  The dashed lines are the result of simultaneous fitting of
  the data for $Q$ = 0 and 1 at $m$ = 0.0016, 0.0048, 0.008
  in the region $7\leq t\leq 13$.
}\label{fig:axialcorr01}
\end{center}
\end{figure}

We now compare the renormalized axial-vector correlation 
function with the QChPT result
\begin{equation}
  \label{eq:AA_QChPT}
  2Z_A^2\sum_{\vec{x}}
  \langle A_0(\vec{x},t)A_0(0,0)\rangle^Q 
  = 2\left(\frac{F_{\pi}^2}{T}+2m\Sigma_{|Q|}(\mu)Th_1(|t/T|)
  \right).
\end{equation}
From the constant term we can determine $F_\pi$, while
$\Sigma$ has to be extracted from the small $t$ dependence.
In Ref.~\citen{Bietenholz:2003bj} it is reported that the
correlators suffer from large statistical fluctuation at
$|Q|=0$ and other topological sectors are insensitive to
$\Sigma$.
We observed similarly large statistical noise as shown in
Figure~\ref{fig:axialcorr01}, but it turned out that 
two-parameter ($F_{\pi}$ and $\Sigma$) fitting does 
work well when we treat the data of different topology and
fermion masses simultaneously. 

As Figure~\ref{fig:axialcorr01} shows, 
our data at $m$ = 0.0016, 0.0048, 0.008 in the 
$|Q|\leq 1$ sectors are well described by the QChPT formula
(\ref{eq:AA_QChPT}).
A simultaneous fit in the range $7\leq t\leq 13$ yields
$F_\pi$ = 98.3(8.3)~MeV and $\Sigma^{1/3}$ = 259(50)~MeV 
with $\chi^2/\mbox{dof}$ = 0.19.
These results are stable under a change of the fit range:
$(F_\pi,\Sigma^{1/3})$ = 
(98.8(8.3)~MeV, 261(47)~MeV) and 
(99.2(8.3)~MeV, 261(45)~MeV) for
$8\leq t\leq 12$ and $9\leq t\leq 11$, respectively.
The result for $F_\pi$ is in agreement with that of the
previous work \cite{Giusti:2004yp}, 102(4)~MeV.
The authors of \citen{Bietenholz:2003bj} quoted a slightly
larger value $\sim$~130~MeV.

\begin{figure}[tbp]
\begin{center}
\includegraphics[width=8cm]{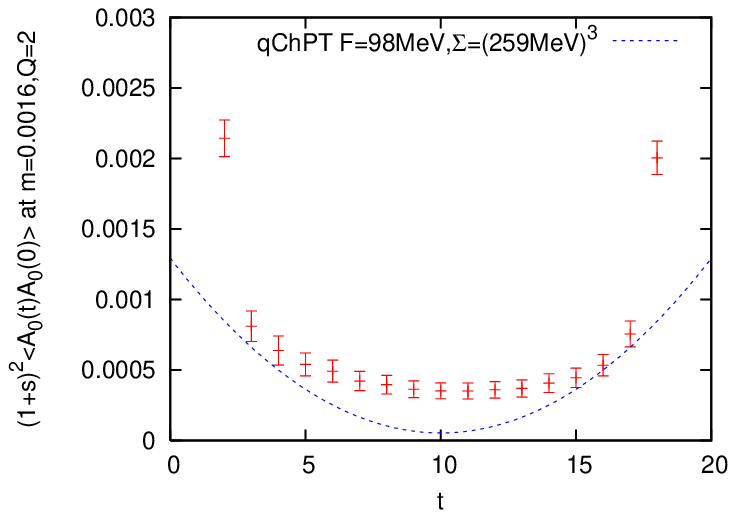}
\includegraphics[width=8cm]{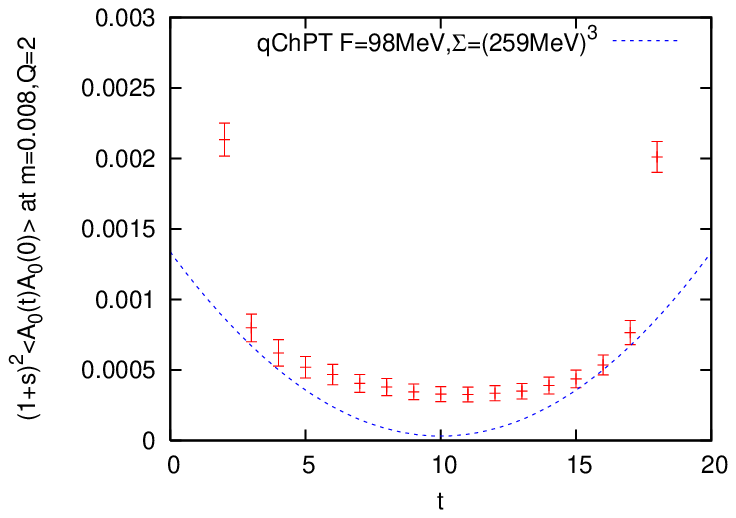}
\caption{
  Axial-vector current correlators at $m$ = 0.0016 (left)
  and 0.008 (right) for $|Q|$ = 2.
  The dashed lines are the QChPT prediction with parameters
  determined through $|Q|$ = 0 and 1 sectors.
}\label{fig:axialcorr2}
\end{center}
\end{figure}

The correlators at $|Q|$ = 2 do not agree well with the
above fit parameters as shown in
Figure~\ref{fig:axialcorr2}.
As discussed in Section \ref{sec:ChPT} it may indicate that
the topological sector $|Q|$ = 2 is already too large to
apply the QChPT in the $\epsilon$-regime.

\subsection{$\Sigma$, $\Sigma_{\mathrm{eff}}$ and $\alpha$
  from connected S and PS correlators}
As discussed in the previous section, the scalar and
pseudo-scalar connected correlators are approximated rather 
precisely only with the lowest 200$+|Q|$ eigen-modes 
at small quark masses ($m$ = 2.6--13~MeV).
In this way we measure the scalar (pseudo-scalar) correlators
\begin{eqnarray}
  \label{eq:connectedcorr_SS}
  \langle \mathcal{S}(t)\rangle^Q
  \equiv
  -2 \sum_{\vec{x}} (1+s)^2
  \langle S^3(x)S^3(0)\rangle^Q_{\mathrm{low-modes}},
  \\
  \label{eq:connectedcorr_PP}
  \langle \mathcal{P}(t)\rangle^Q
  \equiv
  2 \sum_{\vec{x}} (1+s)^2 
  \langle P^3(x)P^3(0)\rangle^Q_{\mathrm{low-modes}},
\end{eqnarray}
for the topological sectors $1\leq|Q|\leq 3$ 
($0\leq|Q|\leq 3$)
at $m$ = 0.0016, 0.0032, 0.0048, 0.0064, and 0.008,
in which the error from higher mode truncation is estimated 
to be only $\lesssim 1$ \%, which can be ignored compared to
statistical errors.
We take an average of the source point over 
$(L/2)^3\times (T/2)$ lattice sites.

To fit our data with the one-loop QChPT formulae
(\ref{eq:SS_QChPTconnected}) and
(\ref{eq:PP_QChPTconnected}), 
we have to determine five parameters in these formulae:
$F_\pi$, $\Sigma$, $\Sigma_{\mathrm{eff}}$, $m_0^2$ and $\alpha$.
Since the sensitivity for $F_\pi$ is weak with these
correlators, we use the jackknife samples of 
$F_\pi$ obtained through
the axial-vector current correlator,
$F_\pi$ = 98.3(8.3)~MeV.
Unfortunately, there still remain too many parameters
to fit with QChPT expressions.
Therefore,
to determine $m_0^2$ we use the relation (\ref{eq:topsus})
and input the value of topological susceptibility
$\chi\equiv\langle Q^2\rangle/V$ from a recent work 
\cite{DelDebbio:2004ns} as
$r_0^4\chi$ = 0.059(3).
It gives $m_0$ = 940(80)(23)~MeV, where the second error 
reflects the error of $r_0^4\chi$.

\begin{figure}[tbp]
\begin{center}
\includegraphics[width=8cm]{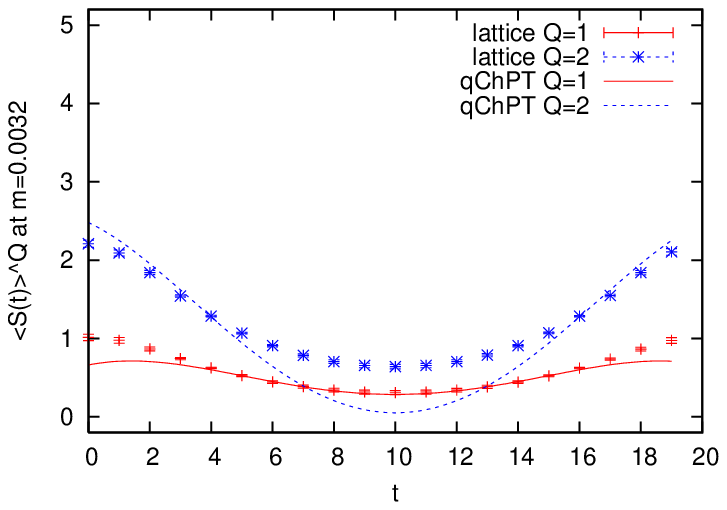}
\includegraphics[width=8cm]{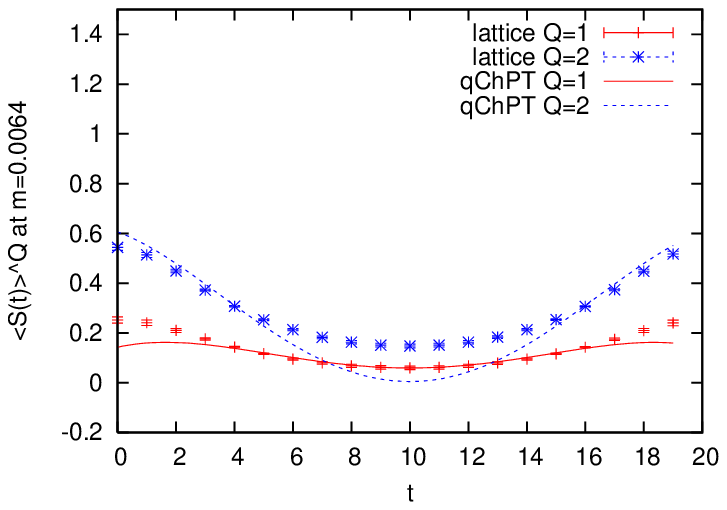}
\includegraphics[width=8cm]{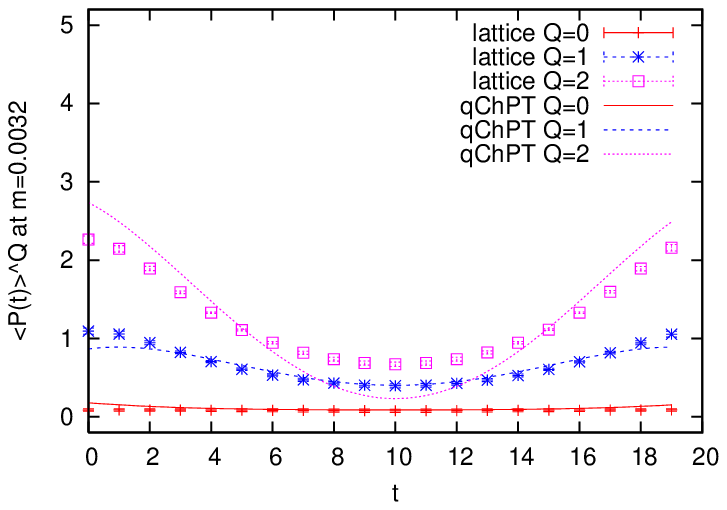}
\includegraphics[width=8cm]{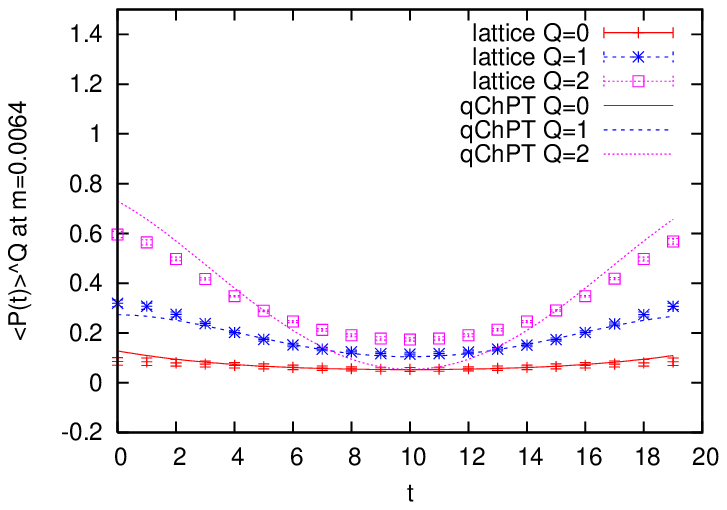}
\caption{
  Scalar (top) and pseudo-scalar (bottom) correlators
  at $m$ = 0.0032 (left) and 0.0064 (right).
  The dotted lines are the fit results with all available
  mass parameters $m$ = 0.0016, 0.0032, 0.0064 and 0.008
  for topological charges 0 and 1.
}\label{fig:connectprop}
\end{center}
\end{figure}

Using these inputs, we fit the correlators (\ref{eq:connectedcorr_SS}) and
(\ref{eq:connectedcorr_PP}) in the range 
$7\leq t\leq 13$ at different $Q$ and $m$ simultaneously.
Figure~\ref {fig:connectprop} shows the correlators with fit
curves.
For $|Q|\leq 1$, the data at all available quark masses
$m$ = 0.0016, 0.0032, 0.0064 and 0.008 are fitted well, and
we actually obtain 
$\chi^2/\mathrm{dof}$ $\sim$ 0.7. (Note that
the correlations between different $t$'s, $m$'s and channels
(PS and S) are not taken into account.)
Our fit results are
$\Sigma^{1/3}$ = 257 $\pm$ 14 $\pm$ 00~MeV,
which is consistent with Ref.~\citen{Bietenholz:2003mi},
$\Sigma_{\mathrm{eff}}^{1/3}$ = 271 $\pm$ 12 $\pm$ 00~MeV, and
$\alpha$ = $-$4.5 $\pm$ 1.2 $\pm$ 0.2, where the first error
is the statistical error and the second one is from uncertainty 
of $\langle Q^2\rangle$. 
They are insensitive to the choice of the fit range.
The central values vary only slightly (for instance, 
$\pm$ 1~MeV for
$\Sigma^{1/3}$ and $\Sigma_{\mathrm{eff}}^{1/3}$) by
choosing shorter fitting ranges $8\leq t\leq 12$ and 
$9\leq t\leq 11$.

From the ratio $\Sigma_{\mathrm{eff}}/\Sigma$ we can
identify the size of the NLO correction in the $\epsilon$
expansion. 
To the one-loop level, it is written as
\begin{equation}
  \frac{\Sigma_{\mathrm{eff}}}{\Sigma} = 
  1+\frac{1}{N_cF_\pi} \left(
    m_0^2 \bar{G}(0) + \alpha \bar{\Delta}(0)
  \right),
\end{equation}
where the parameters $\bar{G}(0)$ and $\bar{\Delta}(0)$ are
regularization dependent.
For this ratio we obtain 1.163(59), which indicates that the
$\epsilon$ expansion is actually converging.

We obtain large negative value for $\alpha$,
which is also reported in Refs.~\citen{Bietenholz:2005kq} and
\citen{Shcheredin:2005ma}. These results contradict
with a previous precise calculation
\cite{Bardeen:2003qz}, which obtained a small value 
$\alpha$ = 0.03(3).
If we instead assume $\alpha$ = 0, and fit $F_\pi$ as a
free parameter, we obtain $F_\pi$ = 136.9(5.3)~MeV and
$\Sigma$ and $\Sigma_{\mathrm{eff}}$ are almost unchanged.
(Detailed numbers are summarized in
Table~\ref{tab:fitresults}.) 
Therefore, there is an apparent inconsistency in the
determination of $F_\pi$ between the axial-vector and
(pseudo-)scalar correlators if $\alpha\sim 0$ is assumed.
A possible cause is that $|Q|=1$ is not small enough to derive
the partition function Eq.(\ref{eq:Z_Q}) (See Appendix).
Eq.(\ref{eq:topsus}) may also have 
a systematic error due to finite $V$ as well as finite $a$. 

The data at higher topological charge, $|Q|$ = 2, are also
plotted in Figure~\ref {fig:connectprop}.
They do not quite agree with expectations from the QChPT
shown by dashed curves in the plots.
A simultaneous fit with all the data including $|Q|$ = 0, 1
and 2 gives a bad $\chi^2/\mathrm{dof}$ ($\simeq$ 12).
This problem of higher topological charge also happens for
the axial-vector correlator as discussed in the previous
subsection.

\subsection{Scalar condensate}

\begin{figure}[tbp]
\begin{center}
\includegraphics[width=8cm]{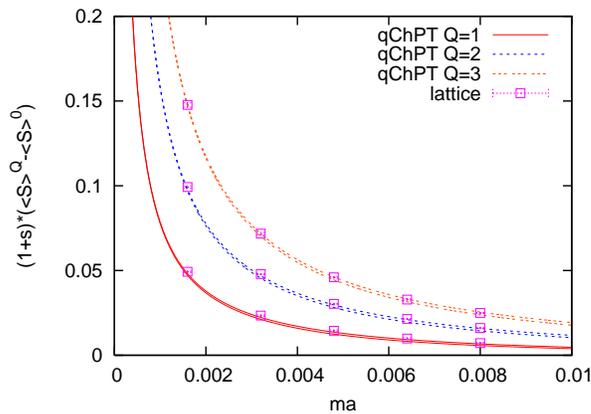}
\caption{
  $-(\langle \bar{\psi}\psi\rangle^Q
  -\langle\bar{\psi}\psi\rangle^0)$
  as a function of quark mass.
  Data points are at $m$ = 0.0016, 0.0032, 0.0048, 0.0064,
  0.008.
  The lines are QChPT predictions with
  $\Sigma_{\mathrm{eff}}^{1/3}$ = 271(12)~MeV.
}\label{fig:scalarconden}
\end{center}
\end{figure}

The free parameter in the scalar condensate is
$\Sigma_{\mathrm{eff}}$ as seen in (\ref{eq:sigmaeff}) and
(\ref{eq:muprime}).
To avoid the problem of ultraviolet divergence we compare a 
difference 
$-(\langle \bar{\psi}\psi\rangle^Q-\langle\bar{\psi}\psi\rangle^0)$
with the QChPT result $\Sigma_Q(\mu')-\Sigma_{Q=0}(\mu')$.
We use the low-mode approximation with 200+$|Q|$ eigenmodes,
and the low-mode averaging is done as in the meson correlators.
Figure~\ref{fig:scalarconden} shows the difference as a
function of quark mass for $|Q|$ = 1, 2 and 3.
We find that the lattice data agree remarkably well with
the QChPT expectation with $\Sigma_{\mathrm{eff}}$ =
271(12)~MeV as determined from the (pseudo-)scalar connected
correlator even in higher topological sectors.
In fact, if we fit the scalar condensate with
$\Sigma_{\mathrm{eff}}$ as a free parameter, we obtain
256(14)~MeV which is consistent with the result above.

\subsection{Disconnected PS correlators}
We also measure the disconnected pseudo-scalar correlators,
which is made possible with the low-mode approximation.
The saturation with 200+$|Q|$ lowest-lying mode is quite
good for the pseudo-scalar channel as discussed in
Section~\ref{sec:low-mode-disconnected}.

In the QChPT it is written as
\begin{eqnarray}
  \label{eq:disconnected}
  \langle \mathcal{P}^d(t)\rangle^Q 
  &\equiv&
  \int\! d^3x\, (1+s)^2 
  \langle 2P^3(x)P^3(0)-P^0(x)P^0(0)\rangle^Q
  \nonumber\\
  & = &
  \int\! d^3x\, (1+s)^2
  \left[
    C_P^d - \frac{\Sigma^2}{2F_{\pi}^2}\left(
      \frac{d_+}{N_c}(m_0^2\bar{G}(x)+\alpha
      \bar{\Delta}(x))
      -e_+\bar{\Delta}(x)
    \right)
  \right],
\end{eqnarray}
where
\begin{eqnarray}
  C_P^d & = & \frac{Q^2}{m^2V^2},
  \\
  d_+ & = & -4\left(1+\frac{Q^2}{\mu^2}\right),
  \\
  e_+ & = & 
  -2\left(\left(\frac{\Sigma_Q(\mu)}{\Sigma}\right)^{\prime}
    -\frac{\Sigma_Q(\mu)}{\mu\Sigma}\right).
\end{eqnarray}

\begin{figure}[tbp]
\begin{center}
\includegraphics[width=8cm]{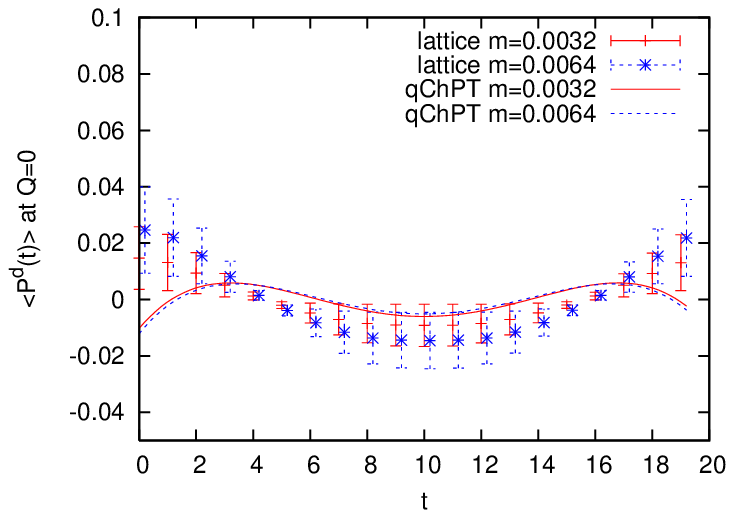}
\includegraphics[width=8cm]{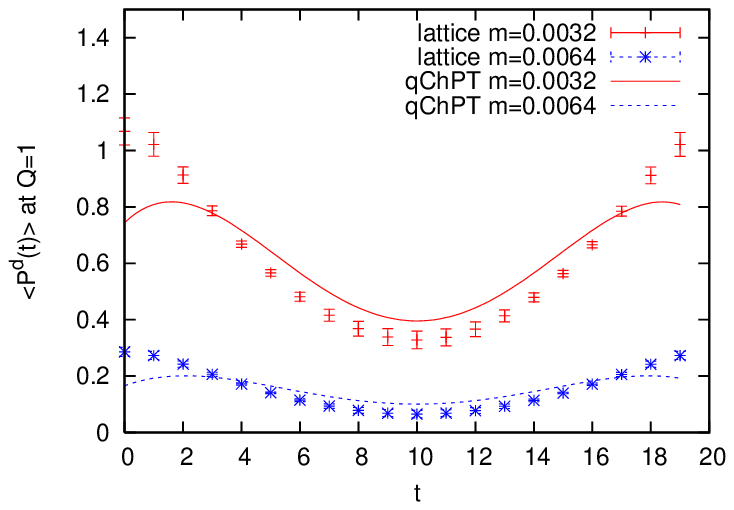}
\includegraphics[width=8cm]{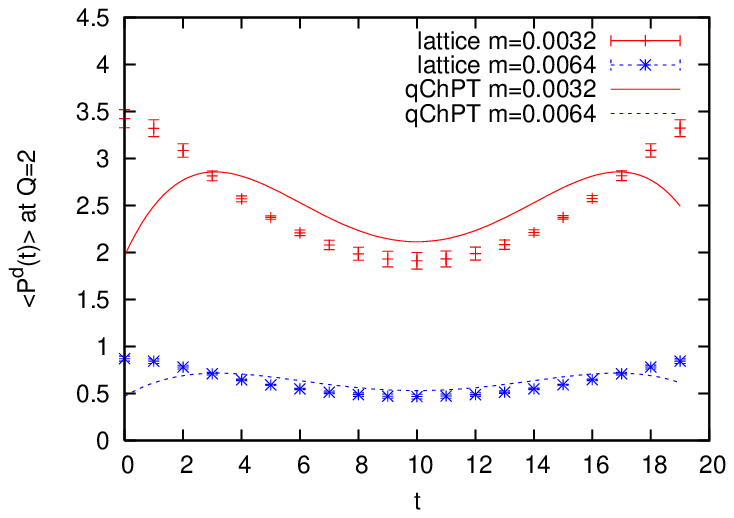}
\includegraphics[width=8cm]{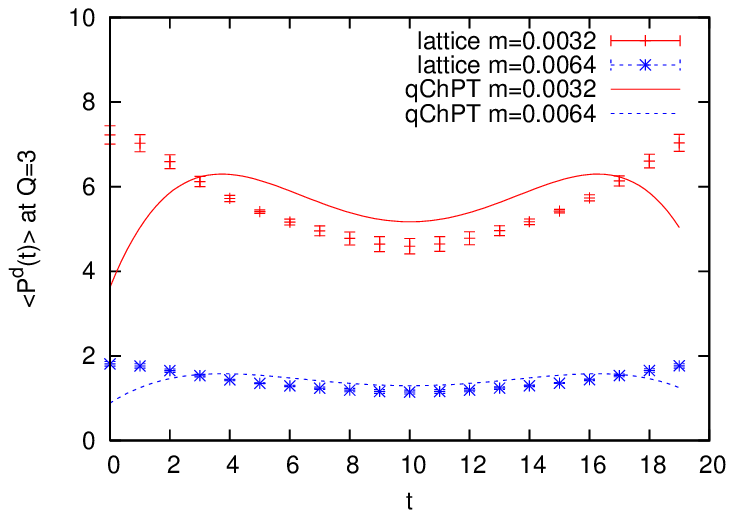}
\caption{
  Disconnected pseudo-scalar correlators in the $0\leq
  |Q|\leq 3$ sectors 
  at $m$ = 0.0032 and 0.0064.
  The curves represent the results of QChPT
  with $\Sigma^{1/3}$ = 257~MeV,
  $F_{\pi}$ = 98.3~MeV, $m_0$ = 940~MeV and 
  $\alpha$ = $-$4.5. 
}\label{fig:disconnectprop}
\end{center}
\end{figure}

In Figure~\ref{fig:disconnectprop} lattice data for
topological sectors $|Q|$ = 0--3 are shown at two
representative quark masses $m$ = 0.0032 and 0.0064.
The QChPT predictions are plotted with the parameters
determined through the axial-vector and (pseudo-)scalar
connected correlators:
$\Sigma^{1/3}$ = 257~MeV,
$F_{\pi}$ = 98.3~MeV, $m_0$ = 940~MeV, and 
$\alpha$ = $-$4.5. 
We observe that the agreement is marginal, though the
correlator's magnitude and shape are qualitatively well
described.
Instead, if we fit the disconnected correlator with
$\Sigma$ and $\alpha$ as free parameters while fixing
$F_{\pi}$ and $m_0$ to the same value, we obtain
$\Sigma^{1/3}$ = 227(32)~MeV and $\alpha$ = $-$3.5(1.2),
which are statistically consistent with those input
numbers. 
Therefore, we conclude that both the connected and
disconnected correlators are consistently described by the
QChPT in the $\epsilon$-regime.
Details of the fit results are listed in
Table~\ref{tab:fitresults}.

\begin{table}[btp]
\begin{center}
\caption{
  Summary of the fitting results.
  The first column denotes the topological sectors used in
  the fit.
  The values in $[\cdots]$ are input parameters.
  The first error is statistical.
  The second and third errors reflect the uncertainty in the
  input parameters,
  $\langle Q^2 \rangle$ and $F_{\pi}$, respectively.
}
\let\tabularsize\footnotesize
\begin{tabular}{ccccccc}
\hline
\hline
correlators
& $F_{\pi}$(MeV) & $\Sigma^{1/3}$ (MeV) &
$\Sigma_{\mathrm{eff}}^{1/3}$ (MeV)
& $\alpha$ & $m_0$ (MeV) & $\chi^2/\mathrm{dof}$ \\
\hline
\multicolumn{7}{l}{axial vector}\\
{\footnotesize $|Q|=0$} & 98(17) & 279(65) &&&&0.02\\
$0\leq |Q|\leq 1$ & 98.3(8.3) & 259(50) & & & & 0.19\\
$0\leq |Q|\leq 2$ & 117.9(4.3) & 335(16) & & & & 2.8 \\
\hline
\multicolumn{7}{l}{connected PS+S}\\
$0\leq |Q|\leq 1$ & [98.3(8.3)] & 257(14)(00)
&271(12)(00) &$-$4.5(1.2)(0.2) &[940(80)(23)] & 0.7\\
$0\leq |Q|\leq 1$ & 136.9(5.3)(0.9) & 250(13)(00)
&258(11)(00) & [0] &[674(26)(16)] & 0.3\\
$0\leq |Q|\leq 2$ & [98.3(8.3)] & 258(12)(00)
&264(11)(00) &$-$3.8(0.5)(0.2) &[940(80)(23)] & 11.8\\
\hline
\multicolumn{7}{l}{disconnected PS}\\
$0\leq |Q|\leq 1$ & [98.3(8.3)] & 227(32)(00)
& &$-$3.5(1.2)(0.3) &[940(80)(23)] & 1.0\\
$0\leq |Q|\leq 1$ & 125.7(5.6)(0.9) & 223(29)(00)
& & [0] &[734(33)(14)] & 0.7\\
$0\leq |Q|\leq 2$ & [98.3(8.3)] & 229(33)(00)(03)
& &$-$3.6(0.2)(0.3)(1.0) &[940(80)(23)] & 1.0\\
$0\leq |Q|\leq 2$ & 135.0(4.9)(1.4) & 237(32)(00)
& & [0] &[684(25)(13)] & 1.7\\
$0\leq |Q|\leq 3$ & [98.3(8.3)] & 229(33)(01)(05)
& &$-$3.6(0.1)(0.2)(0.8) &[940(80)(23)] & 1.0\\
$0\leq |Q|\leq 3$ & 139.3(4.1)(1.4) & 244(32)(00)
& & [0] &[663(19)(12)] & 1.9\\
\hline
\multicolumn{7}{l}{scalar condensate 
($\langle \bar{\psi}\psi\rangle^0-\langle \bar{\psi}\psi\rangle^Q$)}\\
$1\leq |Q|\leq 3$ & & & 256(14) & & &1.2\\
\hline
\end{tabular}\label{tab:fitresults}
\end{center}
\end{table}

\section{Conclusions}
\label{sec:conclusion}

In the $\epsilon$-regime of ChPT, the meson correlators
are largely affected by the fermion zero-mode, and thus by
the topological charge of background gauge field.
This expectation from the effective field theory is
explicitly confirmed by a first-principles calculation of
lattice QCD using the overlap Dirac operator, with
which we can preserve the exact chiral symmetry at finite
lattice spacing.

To reach the $\epsilon$-regime we need small quark masses to
satisfy the condition $M_\pi L\lesssim 1$.
This can be achieved by using the eigenmode decomposition of
the fermion propagator.
Then, the connected scalar and pseudo-scalar correlators are
precisely reproduced by using only 200 low-lying
eigenmodes.
This number would be unchanged even when we decrease the
lattice spacing, as far as the physical volume is kept fixed
to $\sim (1.2\mbox{~fm})^3\times (2.4\mbox{~fm})$, since the
small eigenvalue distribution depends only on a combination
$\lambda\Sigma V$ in the $\epsilon$-regime.
The $a\to0$ limit affects higher modes only, which are 
irrelevant to the low energy dynamics.
For those connected meson correlators the small quark mass
regimes are reached without extra computational costs, and
we can also employ the low-mode averaging technique to
substantially reduce the statistical noise due to near-zero
mode contributions. 
For the axial-vector current correlator, on the other hand,
the saturation by low-lying modes is much worse and we had
to treat them exactly using the (costly) CG solver.
We also investigate the disconnected pseudo-scalar
correlator using the low-lying mode approximation.
The disconnected diagrams are usually very expensive as they
need many fermion inversions, but with this approximation
they are obtained without extra costs.
We confirm that the disconnected pseudo-scalar correlator is
well saturated by 200 low-lying eigenmodes for our lattice.

Remarkable $Q$ and $\mu$ dependences of the quenched ChPT
are well reproduced by our lattice calculation
at $\beta$ = 5.85 on a 10$^3\times$20 lattice.
Then we are able to extract some of the low energy
constants:
$F_\pi$, $\Sigma$, $m_0^2$ and $\alpha$.
The last two describe the artifact of the quenched
approximation.
Fitting our data for meson correlators simultaneously at
different quark masses and topological charges,
we obtain 
$F_\pi$ = 98.3(8.3)~MeV, 
$\Sigma^{1/3}$ = 257(14)(00)~MeV
($\Sigma_{\mathrm{eff}}^{1/3}$ = 271(12)(00)~MeV),
$m_0$ = 940(80)(23)~MeV, and $\alpha$ = $-$4.5(1.2)(0.2), 
from the connected correlators
with $|Q| \leq 1$.
In these numerical results the second error reflects the
error of $\langle Q^2 \rangle$.
We also obtain consistent results from disconnected
pseudo-scalar correlator and the chiral condensate.

Despite such remarkable success of QChPT, we also find
problems.
First, the correlators in $|Q|\geq 2$ sectors are not well
fitted by the parameters determined at $|Q|\leq 1$.
This may indicate a problem of the $\epsilon$-expansion in
QChPT, which comes from the fact that the partition function
at a fixed topology can be justified only for 
$|Q|\ll \langle Q^2\rangle$.
Since $\langle Q^2\rangle$ is proportional to $V$, this
condition is relaxed on larger lattices.
It is therefore interesting to extend our study to larger
volumes.
Secondly, the numerical result for $\alpha$ is rather large
and negative.
Such a large value may pose a question on the validity 
of the partition function (\ref{eq:Z_Q}).
Previous lattice calculations, {\it e.g.}
\cite{Bardeen:2003qz}, indicated the value consistent with
zero and our result is clearly inconsistent with them.
If we assume $\alpha$ = 0, then our data prefer larger value
of $F_\pi\simeq$ 130~MeV, which contradicts with the result
of the axial-vector correlator.
Our calculations are, of course, not free from other
systematic errors due to higher order terms in QChPT, 
finite lattice spacing, {\it etc}.
However, because of the exact chiral symmetry the finite
lattice spacing error does not spoil the consistency with
(Q)ChPT, while the extracted parameters are contaminated.

Once these problems are (positively) solved, the lattice
simulation in the $\epsilon$-regime could become a strong
alternative to the conventional large volume (or large quark
mass) simulations.
A clear advantage is that the small enough quark masses can
be reached and there is practically no question on the
applicability of ChPT.
Obviously, the dynamical simulations in the $\epsilon$-regime
are most desirable.
With the overlap fermion we do not see any fundamental
problem, since the overlap-Dirac operator is
well-conditioned even in the massless limit, though
it is numerically too expensive on the current-generation
machines.

\section*{Acknowledgments}\label{sec:acknowlegments}
We thank Hideo~Matsufuru, Tetsuya~Onogi, and Takashi~Umeda 
for useful discussions and comments.
We also thank Pilar Hern\'andez and Shinsuke~M.~Nishigaki
for valuable discussions.
Numerical works were done at various places:
Alpha workstations at YITP,
Itanium2 workstations at KEK,
and NEC SX-5 at 
Research Center for Nuclear Physics, Osaka University.

\appendix
\section*{}
\label{sec:Appendix}
In this appendix we derive the partition function at a fixed
topological charge (\ref{eq:Z_Q}).
It is obtained from the partition function
(\ref{eq:fullpart}) by Fourier transforming in $\theta$.
\begin{eqnarray}
  \label{eq:ZQinduce}
  Z_{Q}(M) &\equiv& 
  \frac{1}{2\pi}\int^{\pi}_{-\pi}\! d\theta e^{i\theta Q}Z(\theta,M)
  \nonumber\\
  &=&
  \frac{1}{2\pi}\int^{\pi}_{-\pi}\! d\theta 
  \int dU^{\prime}_0d\xi (\mbox{Sdet}U_0^{\prime})^Q
  \exp \left[-\int d^4x\left(
      \mathcal{L}+i\frac{\sqrt{2}Q}{F_{\pi}V}\Phi_0\right)\right]
  \nonumber\\
  &=&
  \frac{1}{2\pi}\int^{\pi}_{-\pi}\! d\theta 
  \int dU^{\prime}_0d\xi (\mbox{Sdet}U_0^{\prime})^Q
  \exp\left[
    -\frac{Vm_0^2}{2N_c}\left(\Phi_0^{\prime}
      -\frac{F_{\pi}\theta}{\sqrt{2}}\right)^2
    -\frac{\sqrt{2}iQ}{F_{\pi}}
    \left(\Phi_0^{\prime}-\frac{F_{\pi}\theta}{\sqrt{2}}\right)
  \right]
  \nonumber\\
  & &
  \times
  \exp \left[
    \frac{m\Sigma  V}{2}\mbox{Str}(U_0^{\prime}+U_0^{\prime -1})
  \right.
  \nonumber\\
  & &
  \left.
    \;\;+\int d^4x\left(
      -\frac{1}{2}\mbox{Str}(\partial_{\mu}\xi\partial_{\mu}\xi)
      -\frac{m_0^2}{2N_c}(\mbox{Str}\xi)^2
      -\frac{\alpha}{2N_c}(\partial_{\mu}\mbox{Str}\xi)^2
    \right)
  \right]
  \nonumber\\
  & = &
  \frac{1}{\sqrt{2\pi\langle Q^2\rangle}}e^{-Q^2/2\langle Q^2\rangle}
  \int dU^{\prime}_0d\xi (\mbox{Sdet}U^{\prime}_0)^Q \exp\left[
    \frac{m\Sigma V}{2}\mbox{Str}(U^{\prime}_0+U^{\prime -1}_0)\right.
  \nonumber\\
  & & 
  +\left.\int d^4x \left(
      -\frac{1}{2}\mbox{Str}(\partial_{\mu}\xi\partial_{\mu}\xi)
      -\frac{m_0^2}{2N_c}(\mbox{Str}\xi)^2
      -\frac{\alpha}{2N_c}(\partial_{\mu}\mbox{Str}\xi)^2
    \right)+ O(\epsilon^4)\right],
\end{eqnarray}
where we use
\begin{eqnarray}
  U & = & U_0e^{i\sqrt{2}\xi/F_{\pi}},
  \\
  \Phi_0 & \equiv &
  \frac{F_{\pi}}{\sqrt{2}}\mbox{Str}(-i\ln U_0),
  \\
  U_0^{\prime} & = & U_{\theta}U_0,
  \\
  e^{iQ\theta} & = &
  (\mbox{Sdet}U_0^{\prime})^Q
  \exp\left(-\int d^4x
    \frac{\sqrt{2}iQ}{F_{\pi}V}\Phi_0\right),
  \\
  \Phi_0^{\prime} & \equiv &
  \frac{F_{\pi}}{\sqrt{2}}\mbox{Str}(-i\ln U_0^{\prime})
  =\Phi_0+\frac{F_{\pi}\theta}{\sqrt{2}}.
\end{eqnarray}
In the last line of (\ref{eq:ZQinduce}), we perform
$\theta$ integral as a Gaussian;
\begin{eqnarray}\label{eq:Qgauss}
  \lefteqn{
  \frac{1}{2\pi}\int^{\pi}_{-\pi} d\theta 
  \exp\left[
    -\frac{Vm_0^2F_{\pi}^2}{4N_c}
    \left(\theta-\frac{\sqrt{2}}{F_{\pi}}\Phi_0^{\prime}\right)^2
    +iQ\left(\theta-\frac{\sqrt{2}}{F_{\pi}}\Phi_0^{\prime}\right)
  \right]
  }
  \nonumber\\
  & = &
  \exp\left(-\frac{Q^2}{2\langle Q^2\rangle}\right)
  \frac{1}{2\pi}\int^{\pi}_{-\pi} d\theta^{\prime} 
  \exp\left[-\frac{\langle Q^2\rangle}{2}\left(
      \theta^{\prime}-\frac{iQ}{\langle Q^2\rangle}
  \right)^2\right]
  \nonumber\\
  & \sim &
  \frac{1}{\sqrt{2\pi\langle Q^2\rangle}}
  \exp\left(-\frac{Q^2}{2\langle Q^2\rangle}\right),
\end{eqnarray}
where $\langle Q^2\rangle=Vm_0^2F_{\pi}^2/2N_c$ and 
$\theta^{\prime}=\theta-\sqrt{2}\Phi^{\prime}_0/F_{\pi}$.
To justify this Gaussian integral, we need a condition
$|Q|/\langle Q^2\rangle \ll 1$,
otherwise the integral
(\ref{eq:Qgauss}) should depend on $\Phi_0^{\prime}$, which
means that $\Phi_0^{\prime}$ and $\theta$ can not be treated
independently and the partition function Eq.(\ref{eq:Z_Q}) is
not valid.
On our lattice with $\langle Q^2\rangle=4.34(22)$,
this condition becomes questionable for $|Q|\neq 0$.


\end{document}